\documentstyle[preprint,aps,epsf]{revtex}

\newcommand{\be}{\begin{equation}}
\newcommand{\bef}{\begin{figure}}
\newcommand{\eef}{\end{figure}}
\newcommand{\hmp}{h^{-1}{\rm Mpc}}
\def\hm{h^{-1}{\rm Mpc}}
\newcommand{\etal}{{\em et al.}}
\newcommand{\ee}{\end{equation}}
\def\spose#1{\hbox to 0pt{#1\hss}}
\def\ltapprox{\mathrel{\spose{\lower 
3pt\hbox{$\mathchar"218$}}
 \raise 2.0pt\hbox{$\mathchar"13C$}}}
\def\gtapprox{\mathrel{\spose{\lower 
3pt\hbox{$\mathchar"218$}}
 \raise 2.0pt\hbox{$\mathchar"13E$}}}
\def\inapprox{\mathrel{\spose{\lower 
3pt\hbox{$\mathchar"218$}}
 \raise 2.0pt\hbox{$\mathchar"232$}}}

\begin{document}
\title{Statistical properties of the LEDA redshift database  }
\maketitle

 
\centerline{}
\centerline{}
\centerline{\Large Luca Amendola}
\centerline{Osservatorio Astronomico di Roma,
V. Parco Mellini 84, 00136, Roma Italy} 
\centerline{\Large Helene Di Nella}
\centerline{ Max planck fur Astronomie, Koenigstuhl 17,
D69117 Heidelberg, Germany}
\centerline{{\Large Marco Montuori} and {\Large Francesco Sylos Labini}}
\centerline{Dipartimento di Fisica, Universit\`a di Roma 
``La Sapienza'' P.le A. Moro 2, I-00185 Roma, Italy.}
\centerline{INFM, Sezione di Roma 1}

\begin{abstract}
We present the statistical properties of large scale galaxy distribution
in the LEDA redshift database. This catalog contains more than 
40,000 redshifts over all the sky. We find that  LEDA, 
although seriously affected
by incompleteness,  shows quite stable 
statistical properties. In particular, we 
have considered the behaviour of the two points correlation function and of 
the power spectrum of the density fluctuations, and we have done several 
tests to check whether  the incompleteness of the catalog
affect these statistical quantities.
Our conclusion is that galaxy distribution in this catalog has fractal
properties up to the distance $\sim 200 \hmp$ with $D = 2.1 \pm 0.2$,
with no sign towards 
homogeneity: this result is statistically stable.
We analyze also the 
properties of the angular distributions, finding a complete agreement 
with the results obtained 
in redshift space. Finally, we compare these results with those 
obtained in other redshift surveys, finding that this
sample well reproduces the properties of galaxy distribution found in the 
different catalogs.

\end{abstract}

\section*{1. Introduction }

In the last twenty years there has been
a very fast development of the observational
techniques in large scale astrophysics.
For example, in 1980 only one thousand redshifts of galaxies were 
known, while now there
 are available about 50,000 redshifts, and in  a few years it is expected that
this number will reach about one million. The possibility of studying
 the large scale 
structure of the Universe by satellites in 
 various electromagnetic bands (from the microwaves to the 
$\gamma$ and X rays)  has improved
 dramatically our knowledge of the Universe. However, despite
 this observational effort, the most popular theoretical models,
 and in particular the Hot Big Bang theory and related
 galaxy formation models, encounter strong difficulties and any 
new observation requires a new "ad hoc" explanation and, 
often, the introduction of new free parameters.

One of the main  tasks of observational Cosmology 
is the identification of the homogeneity scale $\lambda_0$
above which the distribution of galaxies may really become
homogenous (we refer the reader 
to Davis $^{1}$ and Pietronero$^{2}$ 
for an up to date discussion of
the two different point
of views on the matter). The present three dimensional data on the
distribution of galaxies already gives us the possibility to study and 
characterize quantitatively the visible matter distribution at least 
up to $150 \div 200 \hmp$. Among the main 
features reported in recent years we cite the large void discovered by
Kirshner and co-workers $^{3}$
and an evidence for cell-like structure of the Universe
reported by Einasto and co-workers $^{4}$.
Subsequently, by using the CfA redshift surveys, 
it has been 
confirmed the existence of voids and discovered a filament in
 the Coma cluster 
region $^{5, 6}$. These observations, as well as many others, 
show that the large scale distribution of visible matter is characterized 
by having strong inhomogeneities of all scales so that the scale of 
the largest inhomogeneities is comparable with the extent of the 
surveys in which they are detected $^{7- 9}$.

In order to study the matter, in this paper
we analyze the statistical properties of the LEDA (Lyon-Meudon 
extragalactic database) redshift catalog $^{10-12}$ .
This database was created in 1983, 
anticipating the growth of the redshift industry of the last decade 
and 
pursuing the philosophy of the RC1 and RC2 catalogs,
and now contains more than 120,000 galaxies with the most 
important
astrophysical parameters: names of galaxies, morphological 
description,
diameters, axis ratios, magnitudes in different colors, radial 
velocities,
21-cm line widths, central velocity dispersions, etc.
In the 12 years since the inception of LEDA, 
more than 75,000 redshifts have been collected, for more than
40,000 galaxies.
In this extended galaxy catalog it is possible to identify
the main morphological features of galaxy distribution in the 
Local Universe, i.e. up to $\sim 200 \hmp$, and in particular 
to study the large scale structure distribution by means 
of a quantitative statistical analysis, as we are going to 
show in the following.  

In LEDA samples, Paturel and co-workers $^{13}$ identified 
a  very large structure that  was called hypergalactic because
it seems to connect several superclusters (Perseus-Pisces, Pavo-
Indus,
Centaurus and The Local Supercluster). 
Di Nella \& Paturel $^{14}$  showed 
that this structure seems to delineate  a 
privileged plane of the Local universe containing 45\% of the
galaxies where only 25\% would be expected if the galaxy distribution was
homogeneous. Here we briefly summarize the main morphological features 
of galaxy distribution in the LEDA catalog.

The usual statistical 
description of galaxy correlation make use
done of the $\xi(r)$ correlation function $^{15}$.
By analyzing the behavior of this function
in the CfA1 redshift survey
Davis \& Peebles  $^{15}$ 
found that galaxy distribution is characterized by having a well defined 
"correlation length" of $r_0 \approx 5 \hmp$. In this interpretation,
this length should be a physical characteristic scale,
and galaxy distribution should be homogeneous at a few times
$r_0$  i.e. $\sim 15 \div 25 \hmp$. However from the simple 
visual inspection
of redshift maps of galaxies  
 one can see that there are inhomogeneities of 
one order of magnitude larger than $r_0$ 
(for example the Great Wall 
has an extension of 
$\sim 170 \hmp$ $^{16}$ ), and hence this statistical analysis
seems to give an inconsistent result. How can a correlation length
of $5 \hmp$  be compatible with the existence of large 
scale structure of $150 \hmp$ ?

In order to clarify this fundamental problem, 
Pietronero and co-workers $^{17-19}$   
 introduced a new statistical analysis that reconciles 
galaxy correlation with the existence of LSS. This
correlation analysis, well known and used in other field of Physics such
as Statistical Mechanics, is appropriate 
for studying highly irregular distributions, 
but, of course, it can be successfully applied also in the case of
smooth (homogenous) distributions.  The new analysis performed by 
$^{19}$   showed that the interpretation of the standard analysis 
performed using 
the $\xi(r)$ two-points correlation function $^{16}$  
is based on the {\it assumption that the sample under 
analysis is homogeneous}. 
In the case in which this
 basic hypothesis is not satisfied the $\xi(r)$ analysis 
gives misleading results because it gives information that 
are related to the sample size rather than to any real physical 
features. For example, Coleman \& Pietronero $^{19}$ found a fractal behaviour in the CfA1 redshift  
survey up to $\sim 20 \hmp$
 with a fractal dimension $D \approx 1.5$. This result implies
that the correlation length $r_0$  defined by $\xi(r_0) =1$
is spurious, and does not correspond to any real feature of
galaxy distribution, but instead it is related to the size of 
the sample analyzed.

 Sylos Labini \etal  $^{20}$ 
found that 
in Perseus-Pisces surveys the fractal behaviour extends
up to $130 \hmp$
with $D \approx 2$. Moreover in Pietronero \etal $^{2}$ (see for a review 
Sylos Labini \etal $^{21}$) we have analyzed the 
behavior of the correlation function 
for all the published redshift surveys (SSRS, CfA, IRAS, Stromlo-APM, 
LCRS, ESP and also some 
cluster catalogs) finding that all the available samples
show fractal correlation with $D\approx 2$ up to limit of the statistical
validity ($\sim 150 \hm$), without any tendency towards homogenization.
In the case of ESP, the fractal behavior extends to $\sim 1000 \hm$;
however, in this case, the analysis
is made difficult because it requires $K$-corrections,
the adoption of a cosmological model, and a modeling of evolutionary
effects.

Here we investigate the correlation properties and the power spectrum
of the redshift samples available in LEDA database. 
By the full correlation analysis of volume limited (hereafter VL) samples,
we are able to extend the behavior of the two points correlation function
up to $\sim 150 \hmp$. This is the only galaxy catalog
currently available
 in which it is 
possible to study the correlation function for 
three decades of distances. We find that,
 in this range of scale, galaxy distribution is
well represented by a fractal distribution with $D \approx 2$.

Moreover we consider the power spectrum (PS)
of the density fluctuations. This is 
one of the most popular and useful ways to
characterize a distribution. In astrophysics the power
spectrum is particularly significant because most models
of primordial galaxy formation  predict a specific spectrum shape;
in the gravitational
linear theory, different $k$-modes evolve independently, so that
the present power spectrum carries direct information
on the 
primordial physical processes that generated the fluctuations.
However even the determination of the PS is based on the assumption
of the existence of a well defined average density.
Currently, the observations allow one 
to compute the {\it luminous matter}
power spectrum on small and 
intermediate scales, say up to $\approx 200\hmp$, 
through the mapping of the local galaxy clustering,
and the {\it gravitational matter}
 power spectrum on large scales ($>1000 \hmp$),
mainly
through investigation of 
the microwave background anisotropies. 
One of the most
extended redshift surveys, the  CfA2 survey  $^{22}$
presents a power spectrum rising with the scale $\lambda=2\pi/k$
at large wavenumbers  as $P\sim k^{-2}$, and  flattening around $150\hmp$.
These features are also confirmed by the SSRS2
 data $^{23}$ .
Other authors find similar trends 
$^{24-27,56}$
although with some differences on the location
of the flattening and on the overall amplitude.

The flattening is usually interpreted
as the convergence to homogeneity of the sample,
reminiscent of the flattening to a vanishing value
 of the correlation function
on scales $>30 \hmp$ $^{28,22}$ 
and of the angular correlation function
at separations $>10^0$  $^{29}$. 
However, all these surveys, when cut
in volume limited subsamples, extend 
to no more than 130 $\hmp$ (in 
CfA2  $^{22}$).
The fact that the flattening occurs just around this scale should 
prompt a more cautious investigation (see remarks
in  da Costa et al.$^{23}$). 

An even stronger caution
should be exerted as one observes that the amplitude of the
power spectra seems to depend 
on the sample size. From CfA101 to CfA130$^{22}$ (PVGH),
for instance, the spectrum amplitude increases by 40\%.
Two  explanations have been proposed so far
for the scaling: luminosity biasing and
sampling
fluctuations. The effect of luminosity biasing, however,
  is found to be insufficient by the same CfA2
authors
(PVGH, see their Fig. 11a) at least on these scales,
and by the Las Campanas authors  $^{56}$.
The second explanation, sampling fluctuations, is unlikely,
in our opinion, for two reasons. First,
if the survey reached the homogeneity scale, the distribution statistics
must approach the Poisson distribution. Then, in a survey
with $N$ galaxies, the sampling
fluctuations go roughly as $\sqrt{N}$: both in the CfA2 and
in the LCRS case, this amounts to
fluctuations of the order of a few percent, certainly not
enough to explain the strong trend in the PS. Secondly, even
if the fluctuations were strong enough, there is no reason
to expect the {\it systematic} scaling reported in CfA2 and LCRS.
Therefore, in the standard context, in which the
inhomogeneities are not supposed to extend to very large scales, such
a scaling is so far unexplained. In
Sylos Labini \& Amendola  $^{30}$ a different explanation, based
on a fractal viewpoint  $^{19}$,  was
instead proposed:
a finite portion of a fractal shows indeed the same kind
of flattening and  scaling as reported in CfA2. 
Both features appear  then a mere 
consequence of the finiteness of the sample.
The scale
of homogeneity, if this interpretation is correct, is then pushed longward of 
$150 \hmp$.

In order to investigate further this important question, a very
large and deep database is needed. In fact,
to minimize the effects of the
window filtering in the PS calculation, a large sky coverage is
necessary.  In this paper we attempt a
determination of the PS in the LEDA
 database extending from a few Megaparsecs
to more than 150 $\hmp$. This is
the most extended galaxy PS so far obtained in literature.
 On the largest scales, the
effects of poor sampling can be significant;
we will be cautious in deriving too strong conclusions from our
data. Nevertheless, the picture we obtain is amazing: no sign
of convergence to homogeneity is reached; no deviation from a
fractal dimension $D\approx 2$ is detected; and, finally, no 
reasonable CDM-like fit seems
possible on our deepest PS.
If these results will be confirmed by more complete surveys, some
radical change in our view of the galaxy clustering should be adopted.

We would like to stress that we have done several tests,
both for the space correlation and for the power spectrum analysis, 
in order to check the stability of the results versus the 
possible incompleteness of the sample (see the remarks in Davis $^{1}$).
 We find that 
this survey, although it is strongly affected by
incompleteness, show a quite stable statistical signal.

A part the direct three dimensional 
correlation analysis, that is clearly 
the most appropriate way to study 
the statistical properties of galaxy distribution, 
it has been claimed $^{31,1}$that 
there are some indirect evidences,
based on the angular properties, that points towards
an homogenous galaxy distribution.
In order to understand the 
properties of the angular distribution, in 
Sylos Labini \etal $^{32}$ we have  
introduced and studied two new concepts, {\em the dilution effect} and 
{\em the small scale fluctuations}, which are essential
for the quantitative analysis of the statistical validity of the 
available galaxy samples.  We show that the various data that are 
considered as pointing to a homogeneous distribution are all
affected by these spurious effects and their interpretation should
 be substantially changed. In this paper we present  also 
several tests
on the angular distributions. 
In particular,  we study the behaviour of the galaxy number counts.
The angular data have clearly less information, 
with respect to the three dimensional ones, 
and must be studied with great caution.
However, we find even in this case 
 a complete agreement with the fractal picture.

\section*{2. The sample}

LEDA is an all-sky catalog, containing more than 40,000
redshifts. The data of the galaxies recorded in the LEDA database
come from many  different references from the 
literature. The major problems concern the reduction of the 
apparent magnitudes of galaxies, as well as other parameters, 
 in an  homogeneous  system,
and the quantification of the incompleteness due to the 
different sampling ratio  in different area of the sky.

Paturel and co-workers $^{10}$ have done a detailed 
study on the apparent magnitudes, and they have given
the formulae to reduce twenty
different magnitudes systems to the  photoelectric $Bt$ system of the Third
Reference Catalogue (RC3). This reduction 
is done in such a way that the 
resulting
magnitudes are completely free of bias or perturbing effects.
Each magnitude is then given with its own error. This error is of course
in the range 0.1 to 1 magnitude depending of the observations and of the
 comparison
between different
 published values for the same galaxy. The major part of the galaxies
have a mean error less than 0.5 magnitude.
The complete list of the references of the
 magnitudes  used in LEDA ia available on the net
$^{11,12}$
  (we refer to this paper also
for more
information of the data contained in LEDA).


The main selection effect that must be studied with great care is
 the incompleteness of the redshift data available in the 
LEDA database for different directions of observation. The 
sampling ratio is different in different angular regions. 

In Fig.\ref{incomp145}
we report the percentage of galaxies 
known in angular catalogs compared to the number of galaxies having
a measured redshift.
 One can see that up to the limiting magnitude $B_t=14.5$, LEDA have a 
rate of redshift measurements of 90\%, if we extend the study
by including fainter galaxies up to a limit of $B_{t}=16$, this
rate falls to 60\%, and with $B_{t} = 17$, LEDA has only
redshift for 50\% of these galaxies. However the absolute sampling ratios,
i.e. the fraction $f$ 
of galaxies with measured redshift versus the total
number of galaxies present in the sky up
to a certain apparent 
magnitude limit (and not only the galaxies with 
measured angular properties) is 
$f \approx 0.9$ for $B_t=14.5$, 
$f \approx 0.5$ for $B_t=16$, and $f\approx 0.1$
for $B_t=17$.

In order 
to investigate the incompleteness of the LEDA database we have 
built two magnitude limited (ML) samples: the first is limited
at $m=14.5$, is complete to 90\%, and it is called LEDA14.5, the second 
is limited at $m=16$ (completeness 50\%, LEDA16). 
Further, to compare the PS with the results from CfA, we  extracted another
 catalog,  cut to reduce it to
the CfA-North boundaries (R.A.
$8^h<\alpha<17^h$, declination $8.5^0<\delta<44.5^0$,
with a further cut to reduce galactic obscuration, see PVGH),
to be denoted as Leda-CfA . 
This sample contains 4593 galaxies in an angular area of 1.23 sr,
amounting to a completeness level of 70\% (infact, CfA2 North contains
in this region 6316 galaxies, see PVGH). 

The problem that we
will discuss in detail is whether the incompleteness in
the deeper samples destroys the statistical relevance of these
samples, or whether it is possible to recover the right statistical 
properties and to control the effects of such incompleteness.
 In particular we will do several quantitative tests to
clarify this crucial point.
In order to avoid the zones of avoidance due to the Milky Way, 
we cut the all-sky samples in the half-skies 
catalogs. In this way we will have for example,
LEDA14.5N that refers to the northern hemisphere, and 
LEDA14.5S that refers to the southern one. 
The samples denominated North and
South in the paper are respectively selected with the galactic latitude:
$b > +10^{\circ}$ and $b < -10^{\circ}$, as usually done in literature.
  We correct the velocity for the heliocentric motions, 
and we use the Local Group reference frame (see for example Park et al.$^{22}$). 
The Hubble constant is $H_0=100 h \cdot km sec^{-1}Mpc^{-1}$.

We have then 5 ML samples: LEDA14.5N and LEDA14.5S, 
LEDA16N and LEDA16S, and LEDA-CfA.
We will analyzed separately each of these samples.
Moreover to perform the space distribution analysis we 
extract various volume limited (VL) samples 
and we report in followings
 their characteristics. In such a way we 
have various independent VL samples, and we will analyze
in the following the statistical properties of all 
these subsamples.

In Tab.\ref{tableda1} we show the characteristics 
of the VL samples extracted from the database limited at
the apparent magnitude $14.5$ (LEDA14.5N), in 
the same sky region of 
CfA1.

In Tab.\ref{tableda3} 
we show the properties of the VL samples
(limited in absolute magnitude)
extracted from the database limited at $14.5$. The parameter {\it p}
is defined in section 4.

Tab.\ref{tableda5} is the same 
as Tab.\ref{tableda3} for the database
limited at the apparent magnitude $16$ (LEDA16N and LEDA16S).

\section*{3. Morphological properties of the spatial distribution
of galaxies}

In the following we describe, how we obtain the 
  largest three dimensional  maps of the 
distribution of galaxies. 
{}From the simple visual inspection of 
angular distributions, 
one can see that
 galaxies are not randomly distributed but that they are
merged in clusters and large structures of several tens of megaparsecs.
It is already known, after de Vaucouleurs studies $^{33, 34}$, that nearby galaxies
form a flat structure, the Local Super cluster, the pole of which 
is defined by galactic coordinates $l=47 \deg$; $b=6 \deg$. 

In 1988, Geller and 
Huchra $^{35}$ 
showed that the Coma filament was indeed a sheet-like structure, the 
so-called {\it great wall}, similar to structures 
predicted by a model due to Zeldovich  $^{36}$. 
In the southern 
hemisphere, another remarkable work was done by 
Da Costa \etal and Fairall \etal $^{37, 38}$
who
exhibited another wall-like structure in the Pavo-Indus region. 
On the other hand, the Perseus-Pisces region was described as a 
chain $^{39}$
or sheet-like structure $^{40}$.

Paturel and co-workers $^{41,13}$ (see Fig.\ref{flam})
discovered  a very large structure that seems to connect 
several superclusters (Perseus-Pisces, Pavo-Indus,
Centaurus and The Local Supercluster, that
was called  hypergalactic plane  (to not refer to the orientation of
the supergalactic system, i.e. our Local Supercluster). 

The pole of this hypergalactic structure was about 
$l=57 \deg$; $b=22\deg$ in galactic longitude and latitude respectively. 

In order to study the three dimensional features of such a structure, 
from the LEDA database, we have 
built a sample of 24,324 galaxies with radial 
velocities smaller than $15,000 km.s^{-1}$.
For  a plane defined by the coordinates of its pole  $lp$ and $bp$
(galactic longitude and latitude, respectively), we counted the number 
of galaxies which are lying in a conic zone of 
 $\pm 15\deg$ on each side of the plane. We computed a systematic
research of the most populated plane, by giving values between 
0 to $90\deg$ for $b$ and between 0 to $360\deg$ for $l$.\\

One can see in the Tab.\ref{Tpole} the result of this 
galaxy count: $d$ is the limiting  distance of the counting (given in
Megaparsec), $Ngal$ is the total number of galaxies within
this limit, $lp$ and  $bp$  are the galactic coordinates which define
the most populated plane and $\% gal$ is the percentage of galaxies
which are lying within $\pm15\deg$ of the previous plane. 
 If the galaxies were uniformly distributed in the
sphere of computation of radius $d$, when looking in a zone of $30\deg$,
i.e. in a solid angle of  $\pi$ steradians, one should find 25\% 
of the total number of galaxies within the sphere.

A position of the pole at 
$lp=52 \deg; bp=16 \deg$ seems particularly stable between 80 Mpc and
200 Mpc.  
This pole is similar to the one given in 1988. It will be used to define
the hypergalactic coordinates hgl, hgb. From the Tab.\ref{Tpole} it is also possible to give a new determination
of the pole of the Local Supercluster plane.
In 1976 de Vaucouleurs gave: $lp=47\deg$ and $bp=6\deg$ for the
galaxies up to about 3,000 $km.s^{-1}$, here we find $lp=46\deg$
and $bp=14\deg$, for the same region, which is in very good agreement.

We show in Fig.\ref{planhyper} 
the map of the three dimensional galaxy distribution in
hypergalactic coordinates. 

The origin of
hypergalactic longitudes is arbitrarily defined as $l(origin)=l(pole)$ 
and $b(origin)=b(pole)-90\deg$. The Cartesian $XYZ$-
coordinates are thus calculated from hypergalactic longitude and
latitude, hgl and hgb respectively, using the following equations:
 \begin{equation}
        X= r\cdot  cos(hgb) \cdot cos(hgl)
 \end{equation}
 \begin{equation}
        Y= r \cdot cos(hgb) \cdot sin(hgl)
 \end{equation}
 \begin{equation}
        Z= r \cdot sin(hgb)
 \end{equation}
where r is the distance deduced from the radial velocity assuming
the Hubble constant to be $H_{o}=75 km.s^{-1}.Mpc^{-1}$. 

One should notice from this  representation  of the galaxy distribution;
\begin{itemize}
\item  Some radial structures at the periphery are very elongated.
These structures are due to the uncertainties on the distance
deduce from the radial velocity.
As a fact the radial velocity measured is not only the cosmological velocity
(as a function of the distance). The radial component of the velocity
 dispersion 
of the galaxies in a cluster is superimposed to this cosmological velocity.
 It then appears a dispersion on the distances, although {\it quasi} no
error exists in direction. Hence the clusters appear elongated along the line
of sight. All these clusters seems to point toward our Galaxy: this
effect is known as the "fingers of god".
Some methods exist to correct for such an observational effect, however in the
present studies we didn't use them, in order to not add any not well
controlled effect.
This implies that one should not give too much signification to
any radial structure which could be due to this artifact.

\item  One can see a zone (vertical on the plot X-Y) where no
galaxy appear. This zone results from the absorption due to
the dust of the disk of our Galaxy, which hide the others
galaxies of the Universe in this direction.
One should hope that the surveys in the infrared with spectroscopic
follow-up will solve this problem.

\item  The most interesting result visible on
Fig.\ref{planhyper} is the empty region surrounding the Local Supercluster
(centered on Virgo cluster),  delimited by a  belt of neighbor Superclusters:
 Perseus-Pisces, Pavo-Indus, Centaurus and the Great Wall.
All these structures seem to be curved towards the direction of the Local
Supercluster  (LSC)
and being located at a mean distance of 70 Mpc from the LSC.
One can't argue that this situation is due to the previous artifact as this
effect is not radial.
A perpendicular projection to the hypergalactic plane
(Fig.\ref{planperpan}) shows that this tendency is three-dimensional
as the Great Wall is actually curved in the direction of the LSC.
\end{itemize}


{}From the previous sample of galaxies limited in apparent magnitude
to  $B_t<$ 14.5,  we constructed volume limited samples with the 
standard procedure $^{16, 19}$. 
We then compute the research for a privileged
plane with this different subsamples. The results are given in
Table.\ref{Tpolevl}.\\
The pole of the Local Supercluster
 (d$<$ 20 Mpc) is found close to the published value by De Vaucouleurs
 in the  RC2 (l=47deg, b=6deg) with 57\% of galaxies
of the sample located in the region of +/- 15 deg of the plane.
On large scale the hypergalactic plane is present with roughly
40\% of the galaxies in the region of +/- 15 deg of the plane.
The value of its pole is very close to the one found with the
galaxy sample limited in apparent magnitude to $B_t < 14.5$.

\section*{4. Three dimensional correlation function}

With the aim of performing an analysis that does not
require any a priori assumption on the nature of the galaxy 
distribution in our samples, 
we have computed the 
conditional density $\Gamma(r)$ $^{19}$.
The function $\Gamma(r)$ measures the conditional density inside 
a spherical shell of thickness $\Delta r$ at distance $r$ from an occupied
point. Having a three dimensional sample with $N$ galaxies, 
one can measure the $\Gamma(r)$
function from all the points of the sample, and then one 
performs the average:
\be
\label{cf1}
\Gamma(r) = \frac{1}{N} \sum_{i=1}^{N} \frac{1}{4 \pi r^2 \Delta r}
\int_{r}^{r+\Delta r} n(\vec{r}_i + \vec{r}) d\vec{r} = <n(r_i+r)>_i
\ee
where $n(\vec{r})$ is the density at $\vec{r}$ and 
the final average  $<...>_i$ refers to all the occupied points $r_i$.
The exponent of $\Gamma(r)$ is therefore a global property of the system and
it is a robust statistical quantity. 
We will prove this
statement with specific tests in the following. 
It is useful to introduce the average conditional density
\be
\label{cf3}
\Gamma^*(r) = \frac{3}{4 \pi r^2} \int_0^r \Gamma(r') 4 \pi r'^2 dr'
\ee
to smooth out the small fluctuations of the $\Gamma(r)$ function.

For a fractal distribution the integrated number of points
in a sphere around an arbitrary occupied point scales as  $^{42}$
\be
\label{cf5}
N(<r) = B r^D
\ee 
where $D$ is the fractal dimension and $B$
is related to the lower cutoffs of the 
structure $^{19, 32}$.
The average density inside a spherical sample of radius $R_s$ is clearly
\be
\label{cf6}
<n> = \frac{3B}{4\pi} R_s^{D-3}
\ee
In this case it is simple 
to show that  $\Gamma(r)$ (and $\Gamma^*(r)$)
will follow a power law behavior as a function of the distance
\be
\label{cf4}
\Gamma(r) =  \frac{BD}{4 \pi} r^{D-3}
\ee
If the system is homogenous the 
conditional density is clearly flat ($D=3$), so that this statistical tool
is the appropriate one 
to study the fractal versus homogeneity properties.
It is simple to show that in the case of a fractal with dimension 
$D$ the conditional average density is 
$\Gamma^*(r) = (3/D) \Gamma(r)$.

It is important to recall here that 
the conditional density is computed in spherical shell, and 
in a certain catalog, the maximum depth up to which it is possible
to compute $\Gamma(r)$ is of the order of the radius of the 
maximum sphere 
fully contained in the sample volume $^{19, 21}$. 
This length is determined by both the depth of the sample and by
its solid angle in the sky.

Before preceding, it is important to remark a fundamental point.
The main point of the LEDA database analysis is that 
from the law of codimension additivity $^{32}$ 
it follows that if
 we cut points randomly from a fractal distribution, its
fractal dimension does not change.
To clarify a point that became essential in the following analysis,
we discuss briefly the law of codimension additivity $^{42, 19}$.
This stases that  the dimension $D_I$ of the intersection of two objects 
with dimension $d_A$ and $d_B$ embedded in a space with dimension $d$ is
\be
\label{cf11}
D_I=d_A+d_B-d
\ee
For example if $d_A = D$ is the fractal dimension of 
galaxy distribution (in the three  dimensional Euclidean space $d=3$), 
and we intersect it 
with random distribution $d_B=3$, i.e. we cut points
in a random manner, we have that 
$D_I=D \; .
$
Hence, if we cut points randomly from a fractal distribution, its 
fractal dimension does not change. The cutting procedure 
can be applied up to eliminate the $\sim 96 - 98 \%$ of the 
points of the structure without changing the genuine properties of
distribution $^{32}$.

{}It is clear that if the
incompleteness effects are randomly distributed in space,
they will not affect the correlation properties of the sample,
unless the percentage of the points present in the sample
is so small that the sampling noise dominates.
On the contrary if there are some systematic effect,
as a poor sampling in certain regions,
the correlation properties can be seriously affected.

To select the {\it statistically fair samples},
i.e. the VL samples that contain enough points to allows 
one to recover the correct statistical properties
of the distribution from which it has been extracted 
(see Sylos Labini \etal $^{32}$), 
we should firstly determine the percentage of the 
total number of points present in these samples
with respect to an ideal complete survey.
This can be done  knowing the value of the lower cut-off $B$
(Eq.(\ref{cf5}))
of the fractal structure. We can compute the fraction 
of the points present in the sample computing the apparent 
lower cut-off $B_a$ 
\be
\label{lc1}
B_a = \frac{4 \pi N_a}{\Omega R^D}
\ee
where $N_a$ is the number of points present in the VL sample,
$\Omega$ is the solid angle of the survey and
$R$ is the depth of the VL sample considered. 
Then the percentage of points is 
\be
\label{cf2}
p= \frac{N_a}{N}=\frac{B_a}{B}
\ee
where $N$ is given by Eq.(\ref{cf5}).
To compute the {\it absolute} lower cut-off $B$, it is possible to
compute the amplitude of $\Gamma(r)$ 
in a VL sample, and then normalize it to the 
luminosity selection effect that sample. We have 
computed it in several VL samples
of Peruses-Pisces and CfA1 $^{32}$ and we have found that 
$B \sim 15 (h^{-1} Mpc)^{-D}$. We will adopt this value in the following 
calculations.
We list in the Tables 1-2 the characteristics of the VL samples
used, with the relative 
percentage. We have checked that if the 
percentage is lower than $\sim 1 \div 2 \%$ 
the correlation properties are destroyed 
(see Sylos Labini \etal $^{32}$).
Hence the VL samples that we use, have the characteristic to 
be {\it statistically fair samples}. The analyses presented here 
concern only the samples for which $p \gtapprox 2 \%$.

The amplitude of $\Gamma(r)$ is related to the lower cut-offs of
the distribution and it  is simply connected to the prefactor
of the average density (Eq.(\ref{cf4})).
To normalize the conditional density in different VL samples
(defined by the absolute magnitude limit $M_{lim}$)
 we divide it by the Luminosity Factor:
\be
\label{cf13}
\Phi(M_{lim}) = \int_{-\infty}^{M_{lim}} \phi(M) dM
\ee
where $\phi(M)$ is the Schechter luminosity function $^{32}$.
As parameter of $\phi(M)$ we adopt $\delta=-1.1$ and $M_*=-19.7$
$^{9}$.
The normalized CF is
\be
\label{cf14}
\Gamma_{n}(r) = \frac{\Gamma_{M_{lim}}(r)}{\Phi(M_{lim})}
\ee
where $\Gamma_{M_{lim}}(r)$ is the usual CF expressed by
Eq.(\ref{cf4}).

In order to control the effect of the incompleteness in the data, 
we have firstly
studied the correlation properties of the two VL
samples extracted from 
LEDA14.5.  These samples are nearly complete, and the
incompleteness is no more than $\sim 10 \%$. To compare our results
with other published we have computed the correlation function in the same
region of the CfA1 redshift survey.
In this case we have a
complete sample because LEDA contains the
CfA1 redshift catalog. We find that the number of points in the
various VL sample of LEDA14.5-CfA1 are 
nearly the same of that of CfA1.
 The conditional density scales as a power law with
 $\gamma=3-D=0.9 \pm 0.2$ ($D = 2.1 \pm 0.2$)
up to the sample limit that is $\sim 20 \hmp$.
The results of our analysis are therefore in perfect
agreement with those of CfA1 $^{19, 21}$.

The next step is to increase the solid of angle of the LEDA14.5N up
to include all the northern hemisphere. 
We find that the
CF function does not
change its slope nor its amplitude,
so that 
 it remains a power law with  the same fractal dimension.
The scaling region extends up to $\sim 30 \hmp$.
Therefore in this way we can control the effect of incompleteness
and conclude that this result confirms
that the small incompleteness
of this sample does not affect the correlation properties of
galaxy distribution.

We have done the same analysis for the VL
samples of LEDA14.5S
finding a result in substantial agreement with that of LEDA14.5N.
This implies that the statistical properties of the LEDA14.5
are robust with respect to the small incompleteness
in the galaxy redshifts that characterize this sample.
The fractal dimension in the VL
 sample of LEDA14.5 turns out to be $D = 2.1 \pm 0.2.$

In Fig.\ref{g1}
there are plotted also the different $\Gamma(r)$ for the various
VL samples of LEDA14.5:
we find that there is a nice matching of the
amplitudes and slopes. This implies that we are computing the correct
galaxy number density in each subsample, and that this average density
does not depend (or it depends very weakly) on the absolute magnitude
of galaxies. 
We have done this kind of analysis for the 
whole north
galactic hemisphere (LEDA14.5N) and for the southern
one (LEDA14.5): in both cases we obtain the same kind of
behaviour for $\Gamma(r)$.

The following step is to analyze the VL samples
of LEDA16  (north and south).
We have studied the behavior of the correlation function $\Gamma(r)$ in the
various VL samples of LEDA16N and LEDA16S.
We find in both case the distribution has long-range correlation
with the same fractal dimension $D = 2.1 \pm 0.2 $, up to $\sim 80-90 \hmp$
(Fig.\ref{gamleda16}).
In some VL samples
of LEDA16 
$\Gamma(r)$ we find indeed a minor deviation from a single power law
that can possibly be attributed to incompleteness.
Such a conclusion is supported also by the Power spectrum analysis (see
Sec.7). 

To compare the amplitude of  $\Gamma(r)$ in the  different samples of
LEDA16, we
consider the normalized CF (i.e divided by the luminosity factor).
We find that the amplitude of the normalized CF is
about the $40 \%$ of that found for LEDA14.5. This is because in this
case the sampling rate is not $90\%$ as it is for LEDA14.5 but it is
about $50\%$ as we have discussed previously. Hence  
a part this factor, we have that
the CF has the same features of that of LEDA14.5.
Even in this case we conclude that
the incompleteness present in this sample is randomly distributed in the sky,
and the correlation properties are only weakly 
affected by it.


We have performed another test in order to check the 
effects of the incompleteness.
We have cut a VL sample in two regions, say one 
up to $ 50 \hmp$ and the other extending from  $50$ to $100 \hmp$. 
We then compute $\Gamma(r)$ in these 
two subsamples, and we find that 
same power law behaviour extending on all scales, up to the 
effective depth of the samples (see Fig.\ref{2n280}).

This test shows that
the correlation properties of the nearby sample are
the same as the deepest one, and that the signal is stable.
This behaviour is  expected in a fractal 
distribution, and is contrary to what one has
 in the case of 
a radial incompleteness, where the nearby points and the 
far away ones have different environments. 
We can conclude again that the incompleteness of the sample 
does not change the genuine behaviour of the conditional average density.

We have performed also another test: we have cut one sample of 
LEDA14.5N and one of LEDA14.5S at the limiting magnitude $M=-19$.
Then we have measured $\Gamma(r)$: we find the same power law 
behaviour in these two
cases, and also the amplitudes match quite well. Then we have 
repeated the same procedure with LEDA16, multiplying
the amplitude of $\Gamma(r)$ by  $0.5/0.9$. In such
 a way we have obtained 4 samples,
all with the same magnitude limit $M=-19$: we find that all the amplitudes 
match quite well, confirming again the statistical stability 
of the catalog.

As long as the
 space and the luminosity density can be considered independent,
 the normalization of $\Gamma(r)$   in different VL
 samples can be simply done by dividing their amplitude by the
 corresponding luminosity factor. 
Of course such a normalization is parametric, because it depends on
 the two parameters of the luminosity function $\delta$ and $M^*$ $^{43}$.
 For a reasonable choice of these two
 parameters we find that the amplitude of the conditional and radial
 density match quite well in different VL samples. We show in 
in Fig.\ref{ledanorm} the results of such normalization.

\section*{5. Standard correlation analysis}

To clarify the basic aspect of the standard 
analysis we have computed the standard two points 
correlation function $\xi(r)$ defined as $^{16}$ 
\be
\label{x1}
\xi(r) = \frac{<n(\vec{r_0})n(\vec{r_0}+\vec{r})>}{<n>^2}-1
\ee
This function is not suitable to study systems characterized by long-range
correlations, as it is widely discussed in Coleman \& Pietronero $^{19}$.
Suppose we have a spherical sample of radius $R_s$ and 
that the distribution is fractal. Then, the conditional density has 
a power law behavior, while it is simple to show that
\be
\label{x2}
\xi(r) = \frac{\Gamma(r)}{<n>}-1= \frac{D}{3} \left(\frac{r}{R_s}\right)^{D-3}-1
\ee
so that its amplitude {\it depends} on the sample size $R_s$
and it is not a power law. In particular the standard definition of
 "the correlation length"  $\xi(r_0)=1$ does not define
a real physically meaningful quantity, but is just
a fraction of the sample size
\be
\label{x3}
r_0 = \left( \frac{D}{3} \right)^{\frac{1}{3-D}} R_s
\ee
Coleman \& Pietronero$^{19}$ have found 
the linear dependence of $r_0$ on $R_s$
in the CfA1 catalog. Here we will check this behavior for 
all the VL samples that we have built. 

In order to measure $\xi(r)$, 
we have adopted the same procedure used for the computation of
$\Gamma(r)$. In fact we have firstly used the VL samples
of LEDA14.5-CfA1, to compare the result with
those of CP92. Then we have extended the analysis to the whole
northern and southern hemispheres.
Applying the same step to LEDA16, we find
that in the deeper VL sample $r_0$ reached the value of
$45 \pm 5 \hmp$ for a value of $R_{eff}  \approx 150 \hmp$
 scaling as a linear function of the sample
size Fig.\ref{xf2}.

This shows  again that the usual $\xi(r)$ is completely misleading
if applied to systems characterized by long-range correlations.
The so-called "correlation length" $r_0$ is just an artifact
due to an inconsistent mathematical analysis. Being only
a fraction of the sample size,  it has no physical meaning.


Let us note that 
$\xi (r)$ is not a power law for a fractal distribution, but it exhibits a
break due to the "-1" term. This means that in a log-log plot there is a 
very sharp
break when $r \rightarrow r_0$
 because beyond this $\xi(r) $ becomes negative.
As a consequence, the function $\xi(r)$ is systematically
steeper that the function $\Gamma(r)$ or $1+\xi(r)$, unless
one fits at very small values of $r$, i.e. at $r\ll r_0$, which
is not usually done.
The fractal dimension is then usually
underestimate  $^{21}$.

\section*{6. Integral from the vertex}


Let us consider now the conditional average as {\it observed from the
 observer}. 
 In order to
discuss this question
 it is important to focus on the role of the {\it  small scale
fluctuations}. We have studied in detail the general problem in Sylos 
Labini et al. $^{32, 21}$  and we refer to these papers for a more complete discussion 
of the matter.

Considering the conditional average as {\it observed from the
 vertex} $n(r)=N(<R)/V(R)$,
 we expect (Fig.3 - insert panel)
 that for very small distances we observe essentially no other
galaxy, and only at
a distance  $\ell_v$ we begin to observe some other galaxy.
For distances somewhat
larger than
$\ell_v$ we expect therefore a raise of the conditional density
because we
are beginning to have some points. It is therefore important to be able to
estimate and control the {\it minimal
statistical length} $\lambda$. In the paper  Sylos Labini \etal $^{32}$
we have discussed in detail the effects of the finite size fluctuations,
and here we present a new argument for the derivation of the length $\lambda$.
At small scale, where there is a small number of points,  there is
an additional
term, due to shot noise, superimposed to the power
law behaviour of $n(r)$, that destroys the genuine correlations
of the system.  Such a fluctuating term can be
cancelled out by making an average over all the points
in the survey.
On the contrary, in the observation from the vertex, only when the
number of points is larger than, say $\sim 30$ then the shot noise
 term can be not important.
 This condition gives
\begin{equation}
\lambda =  5 \left( \frac{4 \pi}{B p \Omega} \right)^{\frac{1}{D}} \approx
\frac{20 \div 60 h^{-1} Mpc}{\Omega^{\frac{1}{D}}}.
\end{equation}
for a typical VL sample with $M_{VL} \approx M^*$, where
 $B$
 corresponds to the amplitude of the conditional density of all
galaxies $^{32, 21}$. This can be estimated from
the amplitude of the available samples
divided by $p$ as
defined in Eq.(\ref{cf4}). This leads to an estimate (for typical catalogues)
$B \approx 10\div 15 (h^{-1} Mpc)^{-D}$ $^{32}$.

The measured
$n(r)$, the radial density computed 
only from {\it the origin}, therefore,
 scales as power law only beyond a certain scale.
At smaller scales,  the sampling fluctuations dominate the behaviour.

In LEDA, the scaling region begins at $\sim 15 \hmp$ and 
extends up to the limit of the sample, as shown in
Fig.\ref{nrleda}.

The sample VL80 shows a fluctuation near the boundary,
and has a $1/r^3$ decay up to $\sim 10 \hmp$ due to 
the depletion of points in this sample $^{21}$.

Clearly, counting galaxies only
from one observer 
induces larger statistical fluctuations than the $\Gamma(r)$ analysis,
where
 one  averages
from {\it each point} of the sample. 

\section*{7. Power spectrum}

Similar results can be obtained by measuring directly the power spectrum,
a statistics which is increasingly popular in cosmology.
Let us recall  the standard PS notation. If we have $N$
galaxies in a survey volume $V$, we can expand
 the galaxy density contrast   in its Fourier components:
$\delta_{\vec{k}} = (\sum_j w_j)^{-1} \sum_{j \epsilon V}
w_j e^{i\vec{k}\vec{r_{j}}} - W(\vec{k})\,,
$
where $w_j$ are statistical weights, and
$~W(\vec{k}) = V^{-1} \int d{\vec{r}} W(\vec{r})
 e^{i\vec{k}\vec{r}}\,~$
is the Fourier transform of the survey window $W(\vec{r})$, defined
  to be unity inside the survey region, and zero outside.
To take into account the angular incompleteness $f(b,l)\in (0,1)$
we take $w_j=1/f(b,l)$.
Of course, this corrects for the known angular
incompleteness,  leaving  any residual radial gradient.
  We apply the standard power spectrum analysis, estimating
 the
noise-subtracted PS
for a strongly peaked window function as (see e.g. PVGH $^{22}$)
\begin{equation}
\label{estim}
 P(\vec{k}) = V_F\left( <| \delta_{\vec{k}}|^2> - \frac{\sum_j w_j^2}
 {(\sum_j w_j)^2} \right)
\left( \sum_{\vec{k}} 
|W_{\vec{k}}|^2 \right) ^{-1} (1-|W_{\vec{k}}|^2)^{-1}\,.\end{equation}
The factor $V_F$, sometimes not explicitly written down because
assumed conventionally to be  unitary,
is the arbitrary cubic volume, larger than
 the survey region, in which the window
 Fourier transform is evaluated.
Notice that Fisher \etal $^{26}$ also
investigate the PS of IRAS at several depths, but they take as
sample density the density at fixed depth; this clearly
renormalizes automatically the power spectrum amplitude, so that
they would not detect any scaling even for a real fractal.
The factor $\:(1-|W_{\vec{k}}|^2)^{-1}$
has been introduced
by Peacock \& Nicholson $^{25}$
as an analytical correction to the erroneous identification
of the sample density with the population density.
For the lowest wavenumbers the power
spectrum estimation (Eq.(\ref{estim})) is not
acceptable, because then the window filter flattens sensibly (see e.g. PVGH).
The PS estimation is then accurate only up to the scale of the sample.
This is shown in Fig.\ref{poisson}, where to check our
 estimator we show the PS for two
Poisson samples of 160 and 360 $\hmp$ with the LEDA geometry
and constant density: the numerical PS is accurate (i.e., constant
and equal to the theoretical value $V/N$) only up to  the survey scale.

In the homogeneous picture, the  spectrum amplitude, just as the
correlation function, is independent of the sample size $R_s$.
The location of the flattening is also a real
scale\--inde\-pendent feature. In the
fractal picture, on the contrary, the PS scales as $R_s^{3-D}$,
and a turnaround of the spectrum occurs
systematically around $R_s$
$^{30}$. For instance, for $D=2$ the analytical
expression for the fractal PS in a sphere of radius $R_s$ is
$P(k)=(4\pi/3)[2+\cos(kR_s)]R_s^{3-D}k^{-D}-4\pi\sin(kR_s)k^{-3}$
$^{30}$. The rapidly-varying functions
of $kR_s$  are in practice averaged out in the angular
average of the spectrum.

The large number of galaxies present in the LEDA database allow us
to disentangle the fractal scaling, which depends uniquely on the
sample size, from the luminosity segregation, which depends on
the average absolute magnitude of the galaxies in the sample.
There is an almost general consensus, in fact,  that intrinsically
brighter galaxies cluster more than fainter ones; however, the
strength of the effect, and the scale to which it extends
are very incertain issues $^{46-50}$
 The difficulty in
reaching more extended and quantitative conclusions lies
certainly in the relative scarcity of measured galaxy redshifts.
For instance, the  CfA2 survey contains only
608 galaxies (North plus South surveys)
 when cut in a volume-limited sample at 130 $\hmp$.
With this small number, it is difficult to investigate the
question of the dependence of the clustering amplitude with 
luminosity.

We begin with Leda14.5N, which is complete to about 90\%, and
consequently relatively free from strong
selection effects.
We plot in Fig.\ref{leda145} the PS for
 the samples as 40,80 and 100 $\hm$
 (here we did not insert the Peacock-Nicholson correction factor,
 to  show more clearly the turnaround scale). The spectra, as well
 as the turnaround scale, shift in
 agreement with the fractal prediction at all scales. This
 first result, which is already in conflict with the standard
 homogeneous view (unless one invokes a strong
 luminosity bias, see below), motivates us to go further.

We go then to Leda-CfA,  which, as we already mentioned,
has the same boundaries and the same magnitude
limit as CfA2 .

We plot in Fig.\ref{psns} the  PS of the VL subsamples of Leda-CfA
at 60, 101 and 130 $\hm$, respectively.
As reported also in PVGH, the spectra amplitudes scale 
systematically with depth. The flattening, too, moves to
larger and
larger scales, as expected in the fractal view.
Notice that if  the scaling is caused by  a {\it constant} bias factor,
we should 
see a parallel scaling of the spectrum, rather than
having the turnaround moving to larger scales.
 In the same figure we compare the spectra at
101 and 130 $\hm$ with the corresponding CfA spectra (PVGH).
The agreement is reasonable, given that Leda is 70\% complete
at this depth. However,  
these volume-limited samples differ in average
magnitude by increments of order $\Delta \hat M\approx 0.5$ so
one may wonder if the luminosity biasing
is responsible of the scaling.
We calculated then the  PS for all the fixed-depth, varying
magnitude subsamples of Leda-CfA (Fig. \ref{ledacfamagn}).

As the figure shows, no clear sign of luminosity biasing can be detected.
Notice also that the increments in magnitude range  from $\Delta
\hat M=1.1$ within Leda-CfA60 subsamples to $\Delta \hat M=0.3$ for Leda-CfA130.
Only the most luminous galaxies, 
those with  $\hat M=-20.8$ in Leda-CfA at 130 $\hm$, have higher amplitudes 
at  scales $<50 \hm$
than fainter samples. 
For as concern Leda-CfA, then, our conclusion is that the luminosity biasing
does not explain neither the PS scale dependence, nor the 
constant scaling, nor the shift of the flattening scale.
All three features, on the contrary, are naturally expected in
a fractal clustering.

Finally we analyse Leda16N.
We plot in Fig. \ref{leda16} the  PS of the three
VL subsamples of Leda16N,
 at 60, 100 and 150 $\hm$. 
The errors, here and in the following plots, are the scatter
among several artificial fractal 
distributions with dimension $D=2$, with the same
geometry  as the real surveys. In samples with $N$ particles,
the error bars have been rescaled
as $\sqrt{N}$.
The same
scaling as in Leda-CfA is found up to the largest scale.
The samples at 40 and 130 $\hm$, not plotted in Fig. \ref{leda16}
for clarity, also follow the same scaling.

  In Fig. \ref{leda16magn}
 we report the PS of all the subsamples at fixed depths, and in Fig.
 \ref{leda16ampmagn} the PS in linear scale for the fixed-depth
 samples at 100 and 150 $\hm$.

Here again, as for Leda-CfA, the trend with average magnitude 
 is neither systematic, nor constant in scale. Only for as
 concerns samples brighter than $M=-20.5$ we notice  
 a definite trend with luminosity.
In Fig. \ref{ave16}, finally, we plot the PS at 60,100 and 150 $\hm$
 averaged among all the subsamples at fixed depth. Overimposed, 
 we plot the least squares fit and the
 fractal theoretical behavior with $D=1.8$.

The agreement is clearly
 excellent.
The fact that the scaling is $k$-independent is important also
to reject the possibility that the scaling is induced by some kind of
redshift distortion (see e.g. Fisher \etal 1993);
 in fact, such an effect would  depend strongly
on wavelength.
  For a detailed discussion
of the luminosity segregation in fractals
we refer the reader to Sylos Labini and Pietronero (1996) $^{60}$.
Multiplying the spectra of the Leda16N samples at$R_s=60,100,
 130, 150 \hm$
samples by the factor  $(150/R_s)^{3-D}$,
with $D=2$,  we reconstruct (see Fig. \ref{all150})
a single power spectrum (to be denoted as Leda-150), valid for
a sample of depth $150\hm$, extending from
a few Mpc to the sample size,
in agreement with the fractal prediction. 

If at  larger scales the
galaxies are homogeneous, our joint Leda-150 PS would represent our best
estimate of the Universe PS; larger samples should then show no amplitude
scaling and their PS should finally decrease to small values, possibly 
matching the COBE
observations.
The joint Leda-150 PS is well fitted by the scale invariant law $k^{-D}$,
with $D\approx 1.8 \pm 0.1$ .
Then we may fit our
Leda-150 PS with a standard CDM fit $P_{cdm}=Ak T^2(k,\Gamma)$, where $A$ and
$\Gamma=\Omega h$ are  free parameters, and $T(k,\Gamma)$ is the Davis
\etal (1985) $^{44}$ transfer function. 
In the case of CfA2 the best fit parameters 
are around $\Gamma\approx 0.23$ and $A\approx   10^6$. In the Leda-150 PS
the  fit with constant relative
errors gives  $\Gamma\approx 0.15$ and $A\approx  10^{7.3} $.

We   performed two more tests on our PS. 
As already mentioned,
Leda14.5 is almost complete,
contrary to Leda16. 
However, the samples at 40 $\hm$ from Leda14.5N and Leda16N
are  consistent within the errors  (see Fig. \ref{compare}).

The same holds, although with a larger scatter, for the samples
at 100 $\hm$.
Further, we cut  the Leda16N sample
at 100 $\hm$ at various values
of $b$. The spectra, plotted in
Fig. \ref{bcut} are all consistent with each other.
If incompleteness  seriously altered our
results, we would have got strong random variations
for the various $b$-cut catalogs, contrary to what we find.
Notice also that smaller samples (higher $b$ cut) have a
smaller turnaround scale.

In summary,
our conclusion based on the power spectrum analysis
is that the fractal model accounts for the PS in the
LEDA database up to 160 $\hmp$. No sign of convergence to homogeneity
is observed. Extending the analysis to $360\hmp$ we find again an impressive
agreement between the fractal prediction and the data, 
both for the North and the South samples, although
we warn against incompleteness effects. 
If our conclusions are confirmed up to
the largest scales, a major change in the interpretation of
the galaxy clustering would  probably be needed.

\section*{8. The power spectrum fit}
 
 So far we did not pay close attention to the slope
 of the PS. We already noticed that all the PS shown in this paper 
 have a behavior $\sim \lambda^{D}$ (i.e. $\sim k^{-D}$), with $D\approx 1.8$,
 as far as the sample finiteness does not distort the spectrum.
 We look now for the dependence of $D$ with sample size and/or
 sample luminosity. We focus here on
  Leda16. Putting $P(k)=AR_s^{3-D} k^{-D}$ we report in Table III
the full set of $A,D$ for the Leda samples we analysed. 
The fit is performed for a sample at limiting depth $R_s$ between the scales
$.1 R_s$ and $R_s$ in order to avoid both the small scale aliasing and the large scale
distortion.
The ensemble average
is $D=1.67\pm 0.17$ (although we 
prefer to quote the value of 1.8 because in closer agreement with
results from Leda-CfA and other data).
 In Fig. \ref{dim} we report the slope $D$ versus 
the average magnitude for all the Leda16 samples. 
A least square fit gives $D(M)=0.008 M+1.5$, indicating that
there is no significant dependence of the fractal
dimension on the absolute magnitude.
In Fig. \ref{fitamp} we report the amplitude value $P_{50}=
P(k=2\pi/50 \hm)$ and $P_{90}=
P(2\pi/90\hm)$ versus $R_s$ (bottom) and versus $\hat M$ (top).
While the dependence on $R_s$ is clear, the dependence on $\hat M$
is not systematic, except for the brightest samples,
which indeed show a rise in power with magnitude, at $\hat M<-20.5$.
In this range, the least square fit gives $\log(P)\sim -0.5 M$.

\section*{9. Comparison with other data}

We can now compare our results for the space correlations analysis 
with those of other redshift surveys. In this respect,
the first element that one has to consider is that 
the correlation function ($\Gamma(r)$ or $\xi(r)$) can be computed
up to a limiting depth $R_s$. Given a certain catalog this depth 
is determined by the maximum depth {\it and the solid angle} (this 
is the reason why one needs to have surveys with large solid angles
and not only deep pencil beams). 

In particular the depth $R_s$ is
 of the order 
of the radius of the maximum 
sphere fully contained in the sample volume $^{19}$.
This means for example that 
in the Las Campanas Redshift Survey $^{52}$ 
($R_{max}=600 \hmp$, $\Omega=0.1 sr$)
$R_s \approx 20 \hmp$. Such a value of $R_s$ corresponds in terms of $r_0$ 
($r_0 = (D/6)^{1/(3-D)} R_s$)
to $\approx 6 \hmp$ (if $D \approx 2$ as it has been measured in the LCRS),
 so that the value found is completely {\it compatible} 
with the fractal picture.

We consider now CfA2: in this case the 
published analysis of CfA2 $^{22}$ refers to a 
region  with $R_{max} \approx 130 \hmp$ and 
$\Omega \approx 1.2 sr$ so that 
in this case $R_s \approx 35 \hmp$ and hence $r_0 \approx 12 \hmp$
as it results from the paper of Park and collaborators.
Note that 
the linear scaling of $r_0$ with sample depth has been found also 
by these authors.
Similar scalings have been obtained in SSRS2 $^{50}$
and the Optical redshift Survey (ORS) $^{53}$.
In the first case, the scaling has been attributed to
luminosity segregation. This conclusion, however,
is based on very sparse sampling (particularly on a volume-limited
sample
at 91 $\hm$ which contains only 67 galaxies) and,
moreover, directly contradicts the fact that in
CfA2, in Las Campanas $^{52}$
and in the same SSRS2 $^{9}$,
the luminosity biasing was found to be insufficient.
In ORS $^{53}$, the luminosity segregation is
explicitely rejected, and the scaling
from 40 to 80 $\hm$ is attributed generically
to large scale inhomogeneities.
Consider now another catalog: Perseus-Pisces. In this case 
in Sylos Labini  \etal $^{20}$  ( see also Guzzo \etal $^{28}$) 
it was found that $R_s \approx 30 \hmp$ and 
$r_0 \approx 11 \hmp$, and a very well defined 
linear scaling of $r_0$ with sample depth (this result
 is implicit in the Guzzo \etal $^{28}$: as far as $\Gamma(r)$ 
has a power law behaviour,
as Guzzo \etal $^{28}$  found up to $\sim 30 \hmp$,
 $r_0$ 
{\it must} be a linear function of
 the sample size, independently from the fact that
the distribution would eventually become
 homogenous at much larger distances). Finally 
in CfA1  $^{19}$ 
a linear scaling of $r_0$ has been found up to the 
value $r_0 \approx 6-7 \hmp$ for $R_s \approx 20 \hmp$.

However there are still some points that seems to 
be compatible with homogeneity:
the IRAS redshift surveys and the behaviour of
angular correlation function as a function of sample depth.
Both these evidences are, in our opinion, 
an artifact due to finite size effects. 
We now discuss briefly these points, 
but we refer to  Sylos Labini \etal $^{32}$
for a more detailed discussion.

Let us start with the IRAS surveys.
Suppose we have a finite portion of
a real fractal, and that we select randomly only a fraction of the points
belonging to the fractal set. It is clear that if the number of points is
 very small,
the underlying
fractal properties cannot be recovered. Rather, the set of points will look
as a very sparse uncorrelated structure. The point we are going to make in
a forthcoming publication is that this sparse fractal set {\it will look like a
(noisy) homogeneous sample}. We can prove this statement by analytical
and numerical techniques. Let us compare the samples we are considering.
 The CfA1 sample has about 1000 galaxies per steradian, the Perseus-Pisces
 (PP) 3000/sr, IRAS 2Jy 200/sr and IRAS1.2 Jy 400/sr (we are comparing
galaxies per steradian since the average depth of these samples is
not very different). For CfA2 this number is much larger  than that of PP.
As one can see, the IRAS samples are by far the most sparse sample: this is the
reason, in our view, why they do not show the expected correlation scaling.
In other words,   the different results of the IRAS samples with respect to
 that of the optical catalogues (CfA1, CfA2, Las Campanas, SSRS2,
ORS, and Perseus
Pisces) can be attributed to systematic effects due to the poor sampling.
Notice that this explanation holds irrespective
of the fractal viewpoint: the IRAS and optical
data are different, and no clear explanation has been offered to
date. Consider also that
the visual impression
confirms that  the IRAS galaxies do {\it not} fill the optical
voids; rather, they belong to the same structure of the optical
galaxies. Hence it is very puzzling how
 the large voids of tens of megaparsecs
seen both in optical and infrared surveys
can be compatible with a very small correlation length ($4 \hmp$
$^{54,55}$). Our
results show very clearly that
the two IRAS samples are not
statistically fair samples because they contain too few points per unit volume.
We added some comments on this in the conclusions.

The problem of the angular correlation function is very close to the problem
of the IRAS data. In the IRAS data the problem is that the survey is too
sparse to see fractality; in angular data the problem is that we are seeing only
one single sky, and thus, even if there is a large number of points
in the survey, we cannot make an average over observers
(the {\it angular} average that is performed in angular correlation function
does not help, because it does not include depth).
 We prove in
a forthcoming paper that the angular correlation function
of a fractal as populated as the real APM survey
 gives the same ``homogeneous'' scaling observed in that survey.
 Were we able to perform an
average over observers, the "scaling"
 would disappear, as expected in a fractal.
Exactly the same argument applies, of course, to the number counts.
The argument of Peebles 
$^{31}$  on the angular correlation function of a fractal
is correct, but incomplete, not taking into account the statistical significance
of real surveys. Note in this respect that one should also explain
the contradiction between the scaling (linear or not) observed in optical
redshift survey, and the scaling observed in angular correlation
functions: the latter
in fact is not compatible with the former, as is clear from the Limber equation.
On the contrary, they are fully 
compatible within  our view.

\section*{10. Discussion and Conclusions}

In this paper we have discussed the statistical properties of
galaxy distribution in the LEDA database. This is the largest
galaxy redshift catalog currently 
available. We have shown that, although this sample
is seriously affected by incompleteness, the 
statistical global properties of the galaxy distribution are quite stable.
We  stress once again that we are detecting a {\it correlated signal}: if 
this were due to the effects of "inhomogeneous holes" in the
selection procedure, then these artificial features
should have very peculiar properties, i.e. they should be fractal with the
same dimension detected in the regions where the 
incompleteness is lower or
 absent. This means that they must be scale invariant in a well 
defined way. But this seems rather peculiar. In fact,
while it is easy, performing a random sampling of a structure characterized by 
long-range correlations (i.e. a fractal),  to destroy the
 correlations because of incompleteness effects,  
 it seems very difficult, on the contrary,
 to create them. 

The result is that the galaxy distribution has a
well defined fractal behaviour 
with $D \approx 2$ up to $\sim 200 \hmp$, that 
is the limit of statistical validity of this sample. No tendency towards
homogenization has been detected at deeper distances.
We have discussed in detail the compatibility of these results 
with those found in different and more complete samples of galaxies
by different authors. The main problem of all these analyses is that
they are based on the assumption that the large scale 
galaxy distribution is homogeneous. In this paper we have tested this
fundamental assumption, finding that galaxy distribution exhibits a highly irregular 
nature characterized by self-similarity. 
The regime of fractal
behaviour is found to extend to more than three decades in the density.
These results can be confirmed or confuted by the forthcoming survey like 
the  Sloan Digital Sky Survey. 

\def\rapj{Ap. J.}
\def\rmnras{MNRAS}
\def\raa{A\& A}

\newpage
\begin{table}
\caption{The VL subsamples of LEDA14.5N-CfA1 ($\Omega =1.8$)
\label{tableda1} }
\begin{tabular}{|c|c|c|c|c|}
\hline
SAMPLE      &   $d_{lim} (h^{-1}Mpc)$ & $M_{lim}$ & $N$&  p       \\
\hline
VL40      & 40 & -18.54     & 445    & 8   \%    \\
VL60      & 60 & -19.43     & 320     & 2   \%    \\
VL80      & 80 & -20.07     & 226     & 1.1 \%    \\
\hline
\end{tabular}
\end{table}
\newpage
\begin{table}
\caption{\label{tableda3} The VL subsamples of LEDA14.5N and LEDA14.5S
(North N=4703, South N=4163)  (cut in absolute magnitude)}
\begin{tabular}{|c|c|c|c|}
\hline
     &      &         &            \\
{\rm Sample} & $d_{lim} (h^{-1}Mpc)$
& $M_{lim}$ & $N$ \\
     &      &         &            \\
\hline
VL18S &  $31.3$ & $-18.00$ & $578$  \\
VL18N &  $31.3$ & $-18.00$ & $1138$  \\
VL185S &  $39.3$ & $-18.50$ & $712$  \\
VL185N &  $39.3$ & $-18.50$ & $1120$  \\
VL19S &  $49.3$ & $-19.00$ & $1034$  \\
VL19N &  $49.3$ & $-19.00$ & $1095$  \\
VL195S&  $61.8$ & $-19.50$ & $1160$  \\
VL195N&  $61.8$ & $-19.50$ & $844$  \\
VL20S&  $77.4$ & $-20.00$ & $675$ \\
VL20N&  $77.4$ & $-20.00$ & $634$ \\
VL205S&  $96.9$ & $-20.50$ & $371$ \\
VL205N&  $96.9$ & $-20.50$ & $282$ \\
     &      &         &            \\
 \hline
\end{tabular}
\end{table}
\newpage
\begin{table}
\caption{\label{tableda5} The VL subsamples of LEDA16N and LEDA16S
(North N=13362, South N=11794)  (cut in absolute magnitude)}
\begin{tabular}{|c|c|c|c|}
\hline
     &      &         &            \\
{\rm Sample} & $d_{lim} (h^{-1}Mpc)$
& $M_{lim}$ & $N$ \\
     &      &         &            \\
\hline
     &      &         &              \\
VL18S &  $61.8$ & $-18.00$ & $4402$  \\
VL18N &  $61.8$ & $-18.00$ & $4011$  \\
VL185S &  $77.4$ & $-18.50$ & $4632$  \\
VL185N &  $77.4$ & $-18.50$ & $4827$  \\
VL19S &  $96.9$ & $-19.00$ & $4481$  \\
VL19N &  $96.9$ & $-19.00$ & $4678$  \\
VL195S&  $121.0$ & $-19.50$ & $3724$  \\
VL195N&  $121.0$ & $-19.50$ & $3761$  \\
VL20S&  $150.9$ & $-20.00$ & $2463$ \\
VL20N&  $150.9$ & $-20.00$ & $2143$ \\
VL205S&  $187.8$ & $-20.50$ & $1305$ \\
VL205N&  $187.8$ & $-20.50$ & $905$ \\
     &      &         &            \\
 \hline
\end{tabular}
\end{table}
\newpage
\begin{table}
\begin{tabular}{|rllllllllll|}
\hline
 $d (Mpc)$   : &  20 &  40 &  60 &  80 & 100 & 120 & 140 & 160 &
180 & 200 \\
 $Ngal$        : & 1037& 2774& 3945& 4939& 5310& 5507& 5593&
5643& 5661& 5683\\
 $lp (\deg)$ : &  46 &  46 &  46 &  51 &  52 &  52 &  52 &  52 &  52
&  52 \\
 $bp (\deg)$ : &  14 &  14 &  17 &  16 &  16 &  16 &  16 &  16 &
16 &  16 \\
 $\% gal $     : &  55 &  46 &  46 &  45 &  45 &  44 &  44 &  44 &
44 &  44 \\
\hline
\end{tabular}
\caption{\label{Tpole}
$Ngal$ is the number of galaxies located at a distance shorter than
$d$ and  $ \%gal$ is the percentage of galaxies located within 
 $\pm 15\deg$ of the plane defined by the pole of coordinates 
 $lp$ et $bp$. This corresponds to the most populated plane within the volume
of Universe of radius $d$.
This percentage is much larger than what expected for an homogeneous 
distribution of galaxies}
\end{table}
\newpage
\begin{table}
\label{Tpolevl}
\begin{tabular}{|rrrrrrrr|}
\hline
VL   &  Angular &  Btlim &  Mlim & number of & lpole & bpole &  \%  of\\
 $d (Mpc)$ & coverage  &   &   &  galaxies &   &  &  galaxies  \\
\hline
 20 & all sky  & 14.5  & -17.0  &  1167 &45  & 14 & 57  \\
 20 & all sky  &  16 & -15.5  & 2145   &43  &  14 & 49 \\
 80 &  all sky & 14.5  & -20.0  & 1112  & 57 & 21 & 47  \\
 80 &  1/2 sky & 14.5  & -20.0  &  555 &  60 & 18 & 47 \\
 80 & all sky  & 16  & -18.5  &8506   & 59 & 12 & 44 \\
 80&  1/2 sky & 16  & -18.5  &  3934  & 58 & 18 &44  \\
 150& all sky  & 16  & -19.9  & 4476  & 59 & 3 & 39  \\
 150& 1/2  sky& 16  & -19.9  &2275   & 59 & 3 & 43  \\
\hline
\end{tabular}
\caption{ Summary table for the research of a privileged plane
within different Volume Limited samples. The pole of the Local Supercluster
 (d$<$ 20 Mpc) is find close to the published value by De Vaucouleurs
 in the  RC2 (l=47deg, b=6deg) with 57\% of galaxies
of the sample located in the region of +/- 15 deg of the plane.
On large scale the hypergalactic plane is present with roughly
40\% of the galaxies in the region of +/- 15 deg of the plane.}
\end{table}
\newpage

\newpage

\bef
\caption{\label{incomp145}
 To investigate  the completeness  of
the database in terms of rate of  measurements
in a survey, we plot the number of galaxies
with a known magnitude (dashed line) and the
number of these galaxies having a measured redshift
in LEDA (continue line). This plot shows the completeness for
LEDA at $B_T<17$
(half-sky). Up to $B_T \sim14.5$, 90\% of the known
galaxies are measured in redshift. After this limit, the database begins to be
incomplete.}
\eef
\begin{figure}
\caption{ \label{flam}
Flamsteed's equal area projection in supergalactic coordinates showing a
structure connecting Perseus-Pisces, Pavos-Indus and Centaurus
Supercluster (see Paturel et al, 1988)}
\end{figure}

\begin{figure}
\caption{\label{planhyper}
Face-on view of the hypergalactic plane, slice of 100 Mpc on the Z axis.
Each point corresponds to a galaxy.
Ours is located at the origins of the coordinates.
We can see that our position is in the outskirts of the Local Supercluster,
which is itself centered on the Virgo cluster.
Some observational 
artifacts are clearly seen: the zones without redshift measurements
due to the absorption of 
the Milky Way (vertical trace) and the radial elongation
of the galaxy clusters (in particular in periphery). The local Supercluster
seems surrounded by an empty region, delimited by the neighbor superclusters : 
Centaurus, Pavo-Indus, Perseus-Pisces and the Great Wall.
} 
\end{figure}

\begin{figure}
 \caption{\label{planperpan}
 Perpendicular view to the hypergalactic plane.
 Slice of 100 Mpc on the X axis. On this view the Great Wall is clearly curved
 in direction of the Local Supercluster. One can also see clusters of galaxies
 elongated radially (see the text for explanation). The
 major part of the galaxies are localized between +50 Mpc and -50 Mpc
 on the Z axis, which correspond to the hypergalactic plane seen by the side.
}
\end{figure}

\bef
\caption{\label{g1} The conditional average density $\Gamma^*(r)$ 
for various VL samples of the LEDA14.5 sample. $N$ for the VL
 sample of the Northern hemisphere and $S$ for the Southern one.
The reference line has a slope $-\gamma =-0.9$ ($D=2.1$).}
\eef

\bef
\caption{\label{gamleda16}  The conditional average density $\Gamma^*(r)$
for various VL samples of the LEDA16 sample. $N$ for the VL
 sample of the Northern hemisphere and $S$ for the Southern one.
The reference line has a slope $-\gamma =-0.9$.}
\eef

\bef
\caption {\label{2n280} The conditional average density $\Gamma^*(r)$
for the VL sample 2N280  LEDA16 sample. 
The filled circles refer to the $\Gamma(r)$ of the 
total sample, the squares  for the first half (0-160 $\hmp$)
and the diamonds for the second half (160-360 $\hmp$).}
\eef

\bef
\caption{\label{ledanorm} 
The spatial density   $\Gamma^*(r)$     computed in some 
VL samples LEDA
and normalized to the corresponding
luminosity factor. } 
\eef 

\bef
\caption{\label{xf2} Behaviour of the so-called "correlation 
length" as a function
of the sample size. The reference line has a linear behaviour
is in  agreement with the fractal prediction. It is remarkable to note that 
in  the deepest sample 
we find $r_0 \approx 45 \pm 5 \hmp$.}
\eef

\bef
\caption{\label{nrleda} 
The spatial density   $n(r)$     computed in some 
VL samples of LEDA145. A very  well defined power law behavior with
$D \approx 2$  is shown for $r \gtapprox 15 \hmp \approx \lambda$.
} 
\eef 

\bef
\caption{\label{poisson} 
Top panel: the PS of two artificial Poisson samples of depth 160 (circles)
 and 360 (squares) $\hm$, with the Leda window geometry.
  Bottom panel: the window PS of the
 two samples.
} 
\eef

\bef
\caption{\label{leda145} 
The PS (without Peacock-Nicholson correction)
of the three VL samples of Leda14.5N at 40, 80, and
100  $\hm$,
from bottom to top, respectively. The  lines are
best fit to the data (at scales smaller than the turnaround).
 The systematic shift in the amplitude, and in the
location of the flattening is clear.}
\eef

\bef
\caption{\label{psns} 
The PS of the  VL samples of 
Leda-CfA at 60,  100, and 130
 $\hm$,
from bottom to top, respectively. The thin line indicates a slope of
$D=2$. The systematic shift in the amplitude, and in the
location of the flattening is clear. 
Also shown are the data from
 CfA2-101 and CfA2-130 PS (from PVGH). }
\eef

\bef
\caption{\label{ledacfamagn} 
Leda-CfA PS at various average magnitudes.
No clear luminosity trend can be detected. }
\eef

\bef
\caption{\label{leda16}
The PS of the  VL samples of Leda16 at 60,  100 and 150
 $\hm$,
from bottom to top, respectively. The top thin line indicates a slope of
$D=2$, the dotted lines are the
best fits to the data. 
For sake of clarity we plot the errorbars only on two sets of data.
The systematic shift in the amplitude is clear. }
\eef

\bef
\caption{\label{leda16magn} 
Leda16 PS at various average magnitudes.}
\eef

\bef
\caption{\label{leda16ampmagn} Leda16 relative amplitudes for
various subsamples in linear scale. Top panel: samples
at 100 $\hm$; bottom panel: samples at 150 $\hm$. The brightest samples 
show
some luminosity dependence.}
\eef

\bef
\caption{\label{ave16}
Average PS of  the equal-depth subsamples of Leda16 at 60, 100
and 150 $\hm$.
The dotted lines are the least square fits; the solid lines
are the fractal predictions with $D=1.8$.}
\eef

\bef
\caption{\label{all150}
The PS of the average samples at 60, 100, 130 and 150 $\hm$ of Leda16
rescaled according to the fractal model with $D=1.8$ to a depth
of 150 $\hm$. The curve is a CDM-like fit with $\Gamma=0.15$.}
\eef

\bef
\caption{\label{compare}
Comparison of spectra at scales of 40 and 100 $\hm$
for samples from Leda14.5 and Leda16 , which have very different
levels of incompleteness. The spectra are consistent
within the typical scatter. The solid lines are the best fits to the
14.5 samples.}
\eef

\bef
\caption{\label{bcut} The PS of the 100 $\hm$ 
VL Leda16N samples cut at $b>20^0,
30^0, 40^0, 50^0$ and $60^0$. }
\eef

\bef
\caption{Fractal dimension \label{dim} versus average absolute magnitude.
The continuous line is the ensemble average dimension, $D=1.67$; the
two dotted lines give the root-mean-square errors.}

\eef

\bef
\caption{The amplitudes $P_{50}, P_{90}$ versus sample depth
(bottom) and sample aver\label{fitamp}age magnitude (top). In the bottom panel,
squares report $P_{50}$, circles report $P_{90}$, and
the dotted line is the fractal scaling trend for $D=1.8$.
In the top panel, open squares, open circles, and filled squares
represent $P_{50}$ for samples at 60, 100 and 150 $\hm$,
respectively; open triangles, crosses and filled triangles
give $P_{90}$ for the same samples. Only the brighter samples
show a definite monotonous increase of power with luminosity.}

\eef

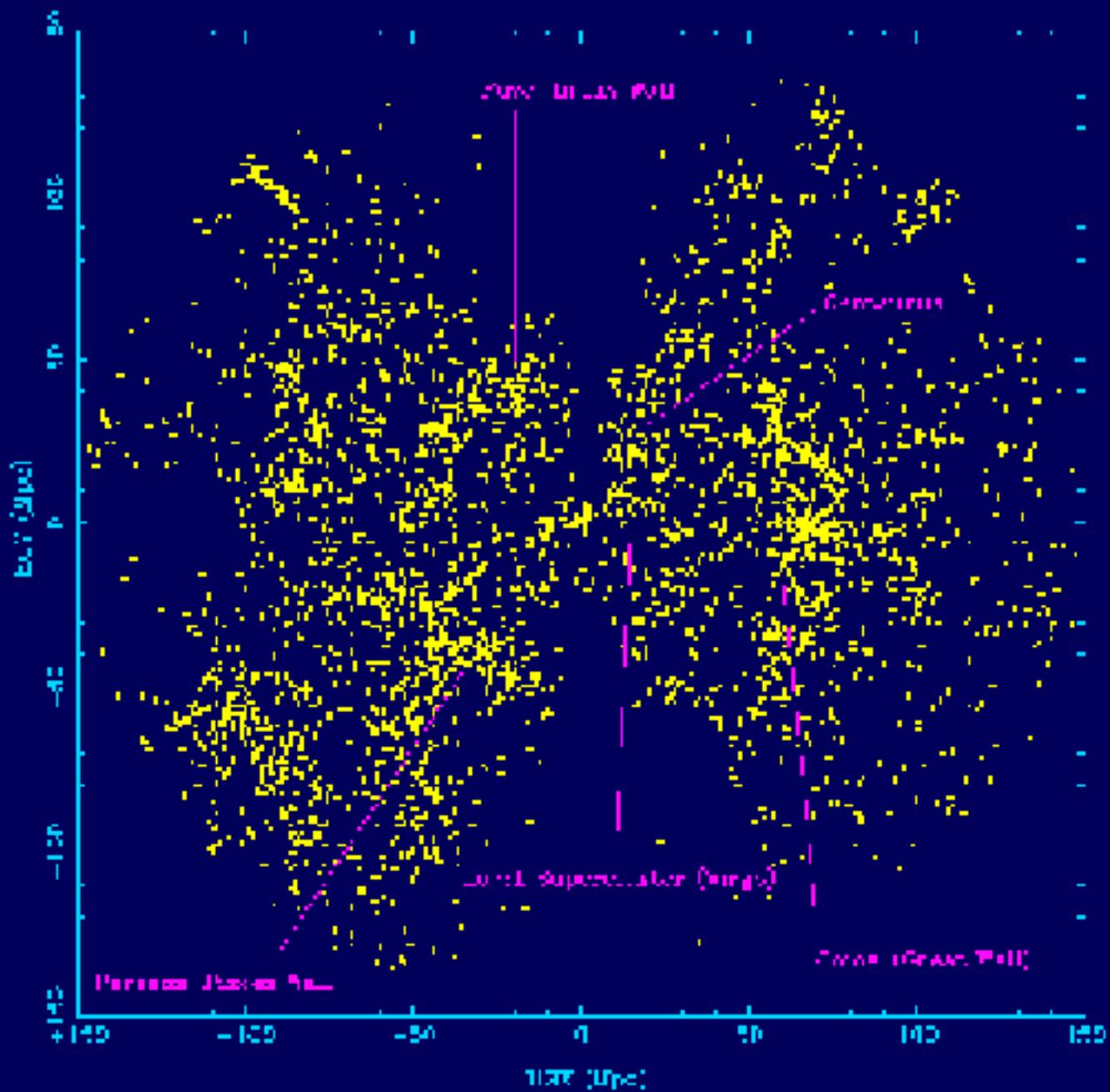
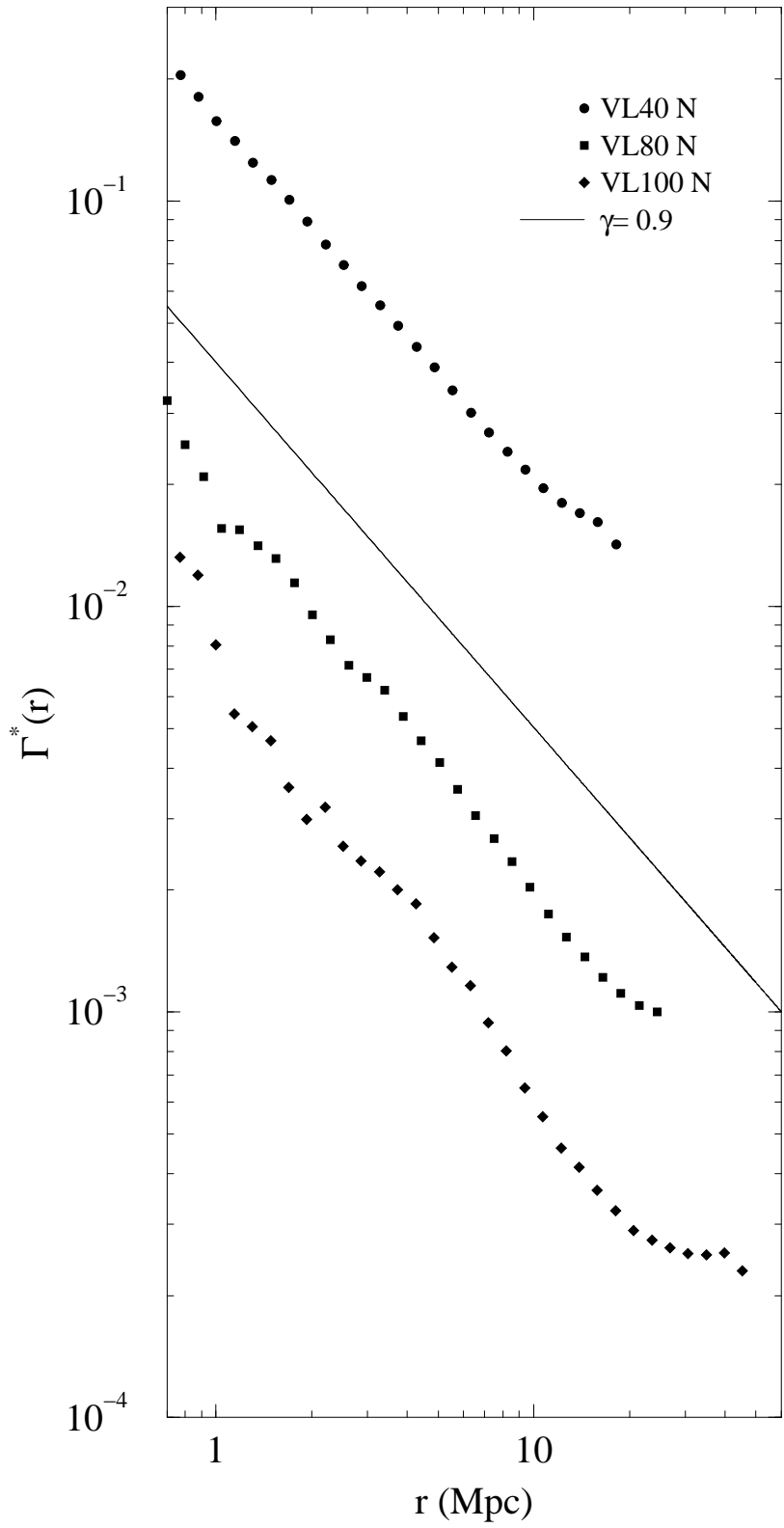
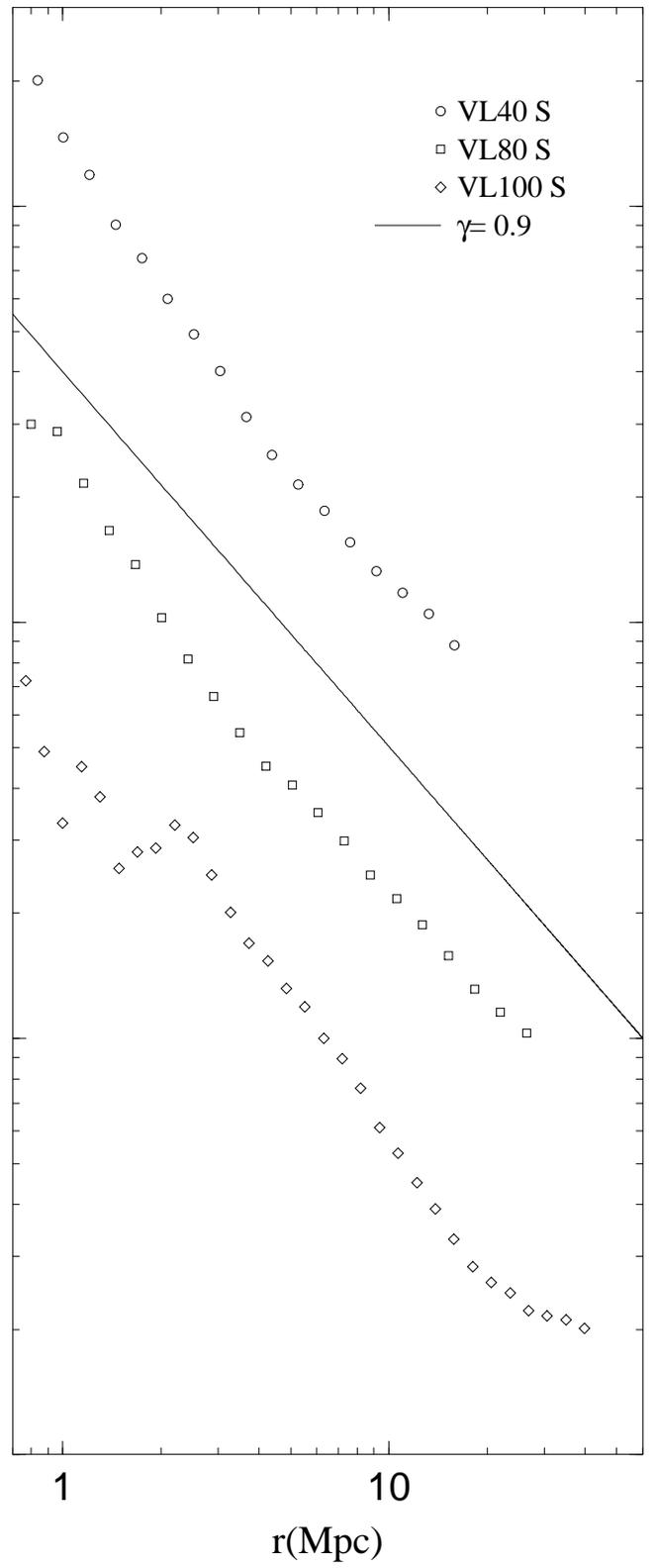
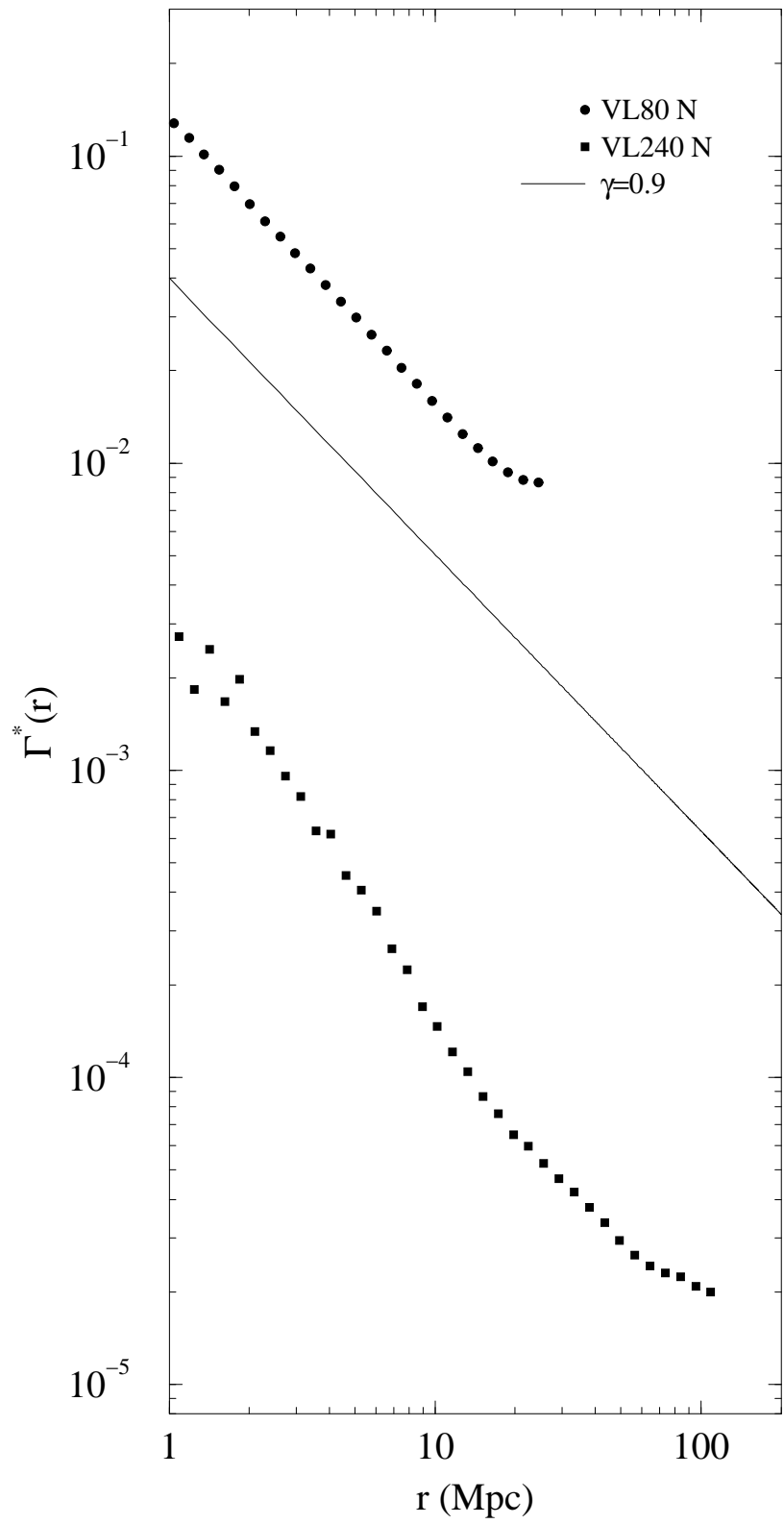
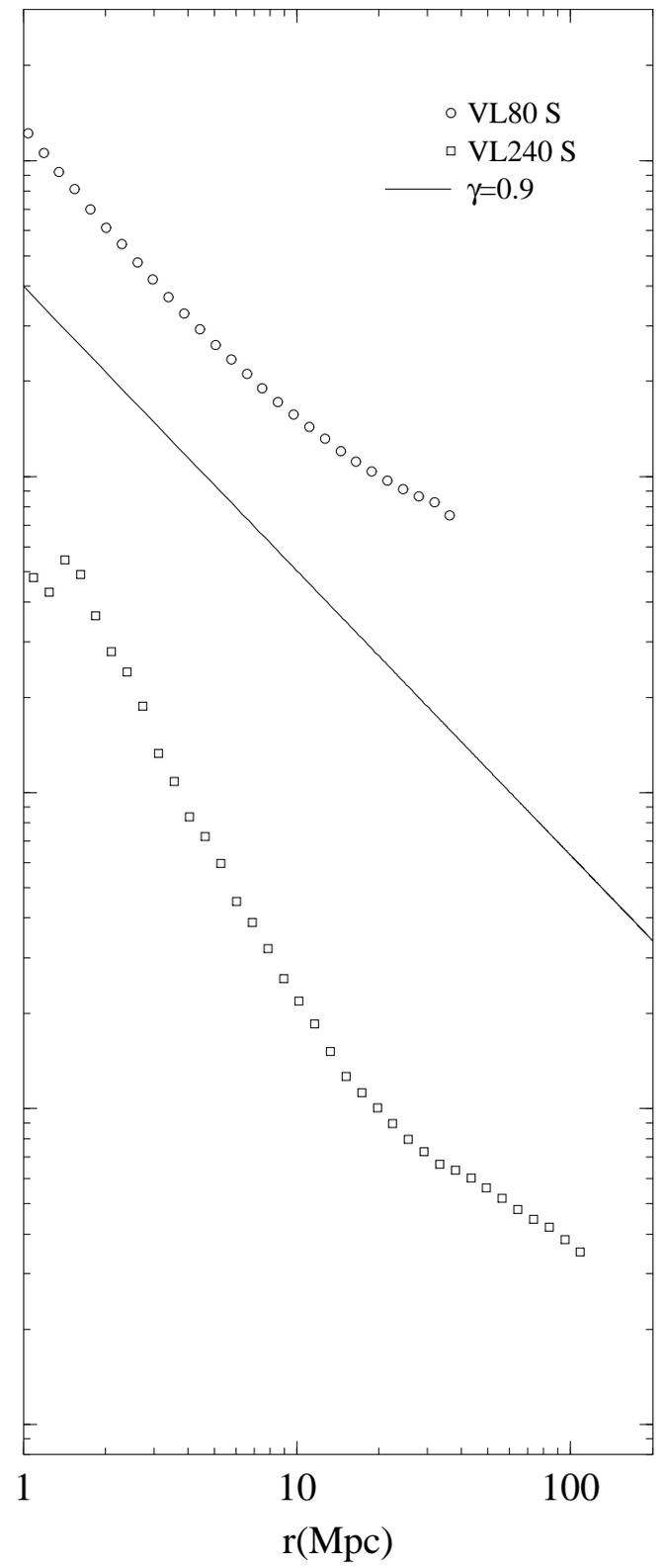
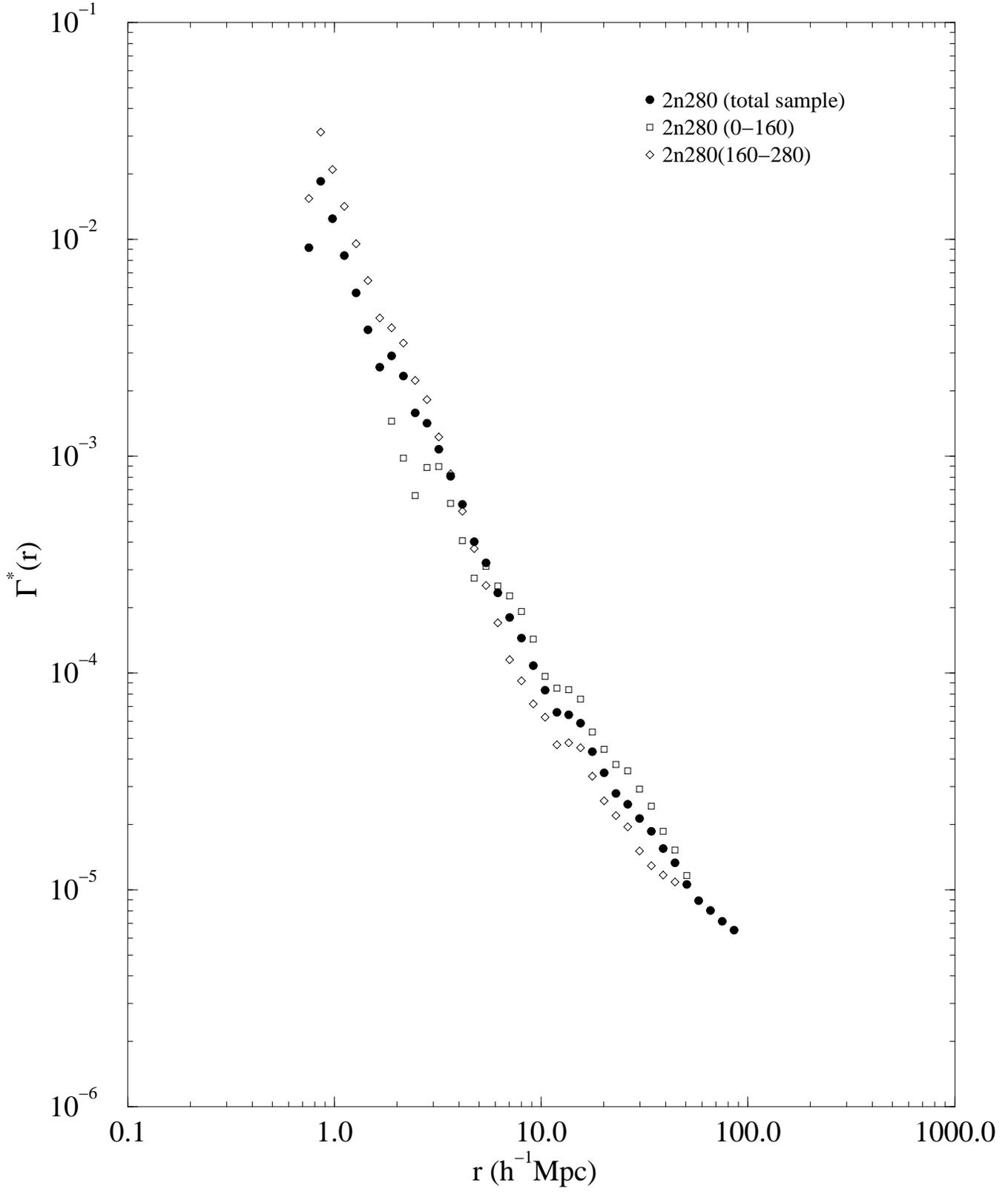
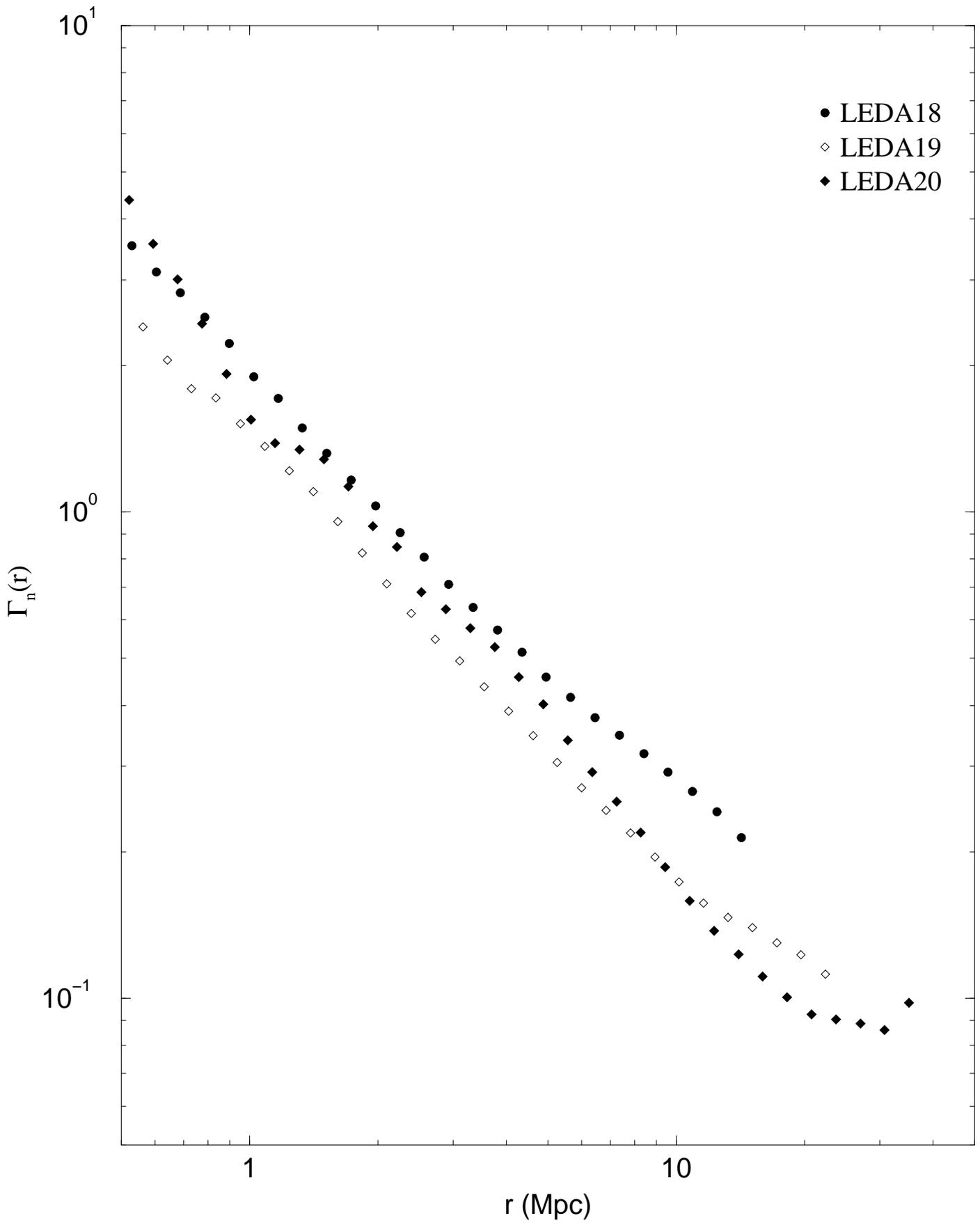
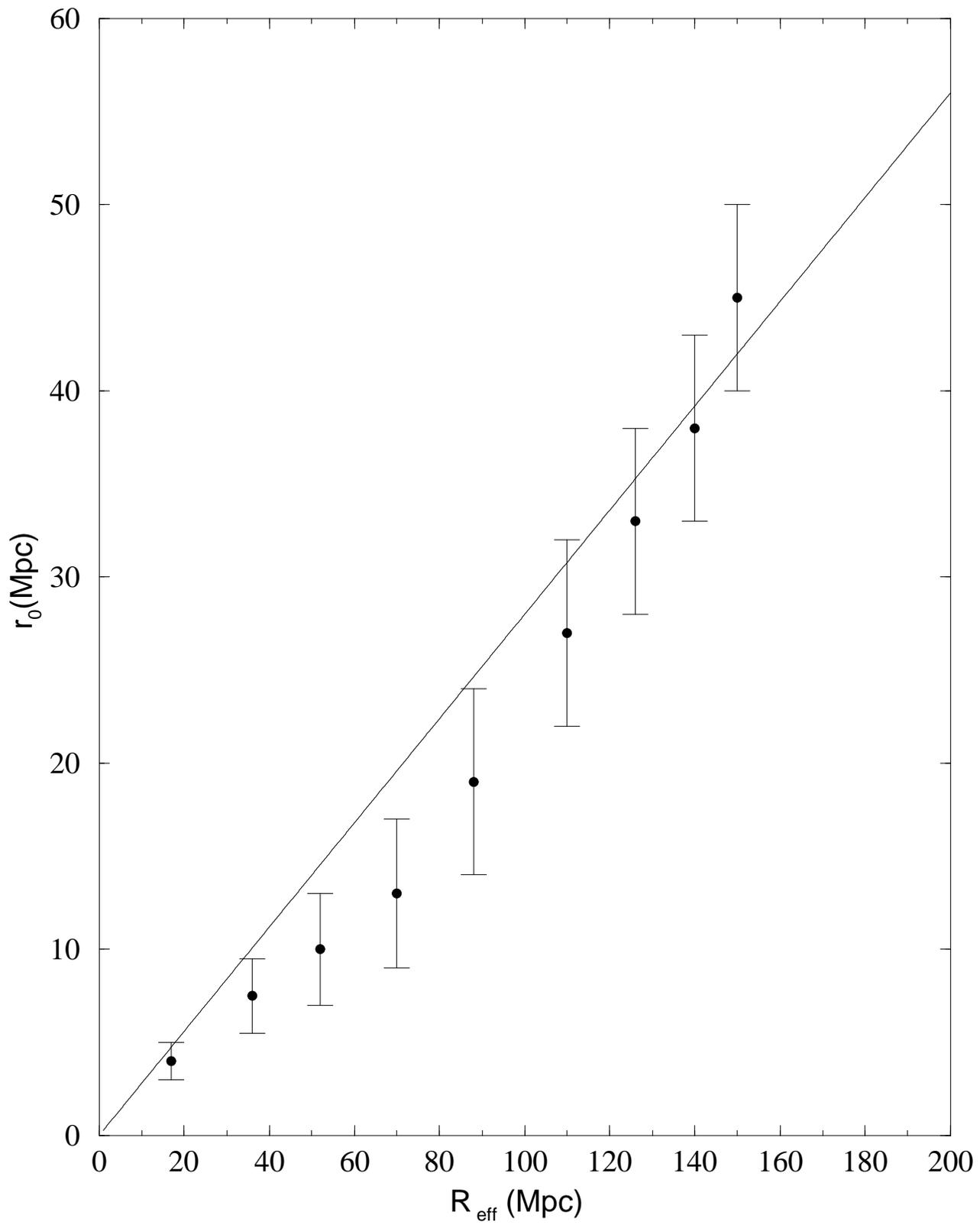
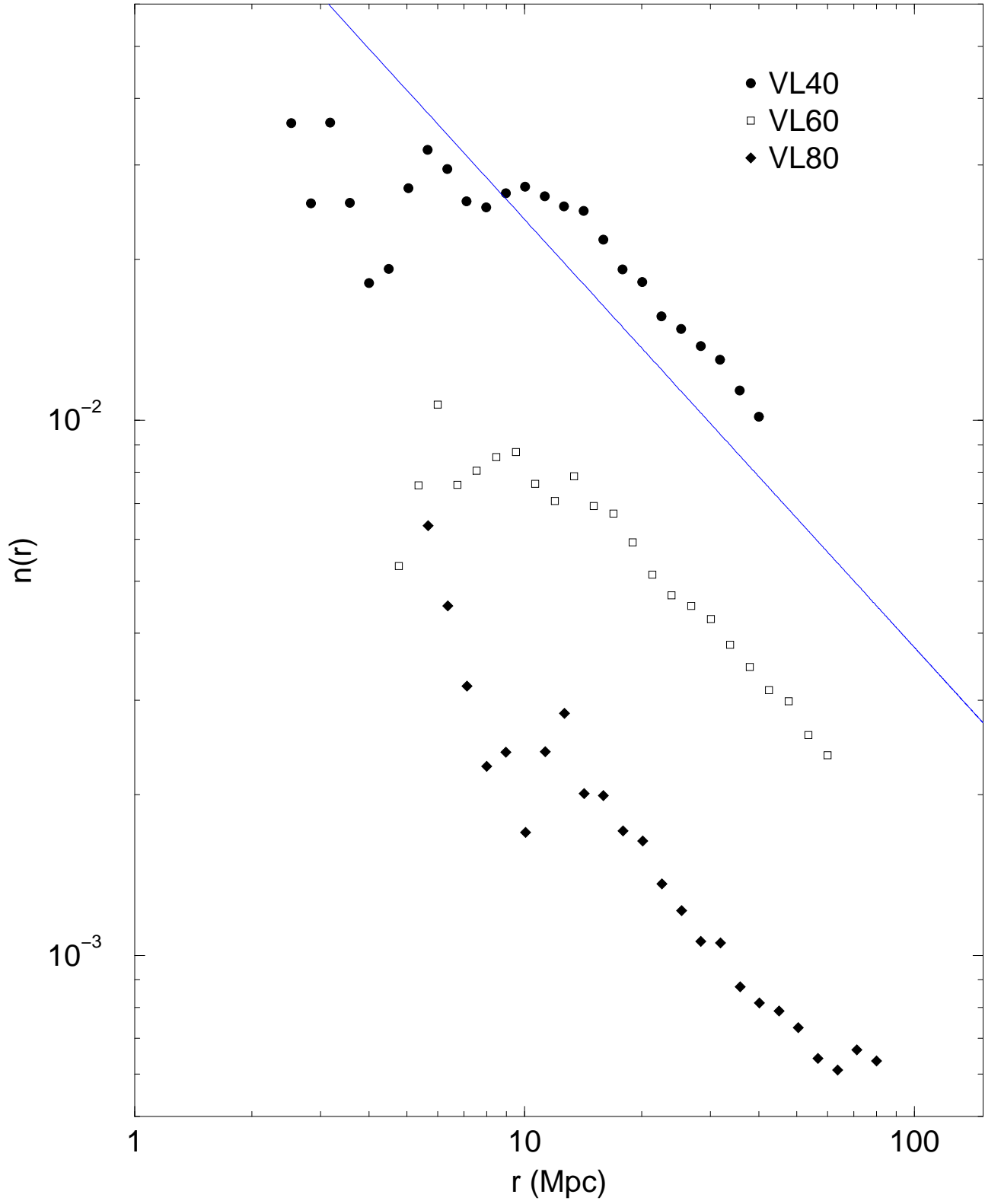
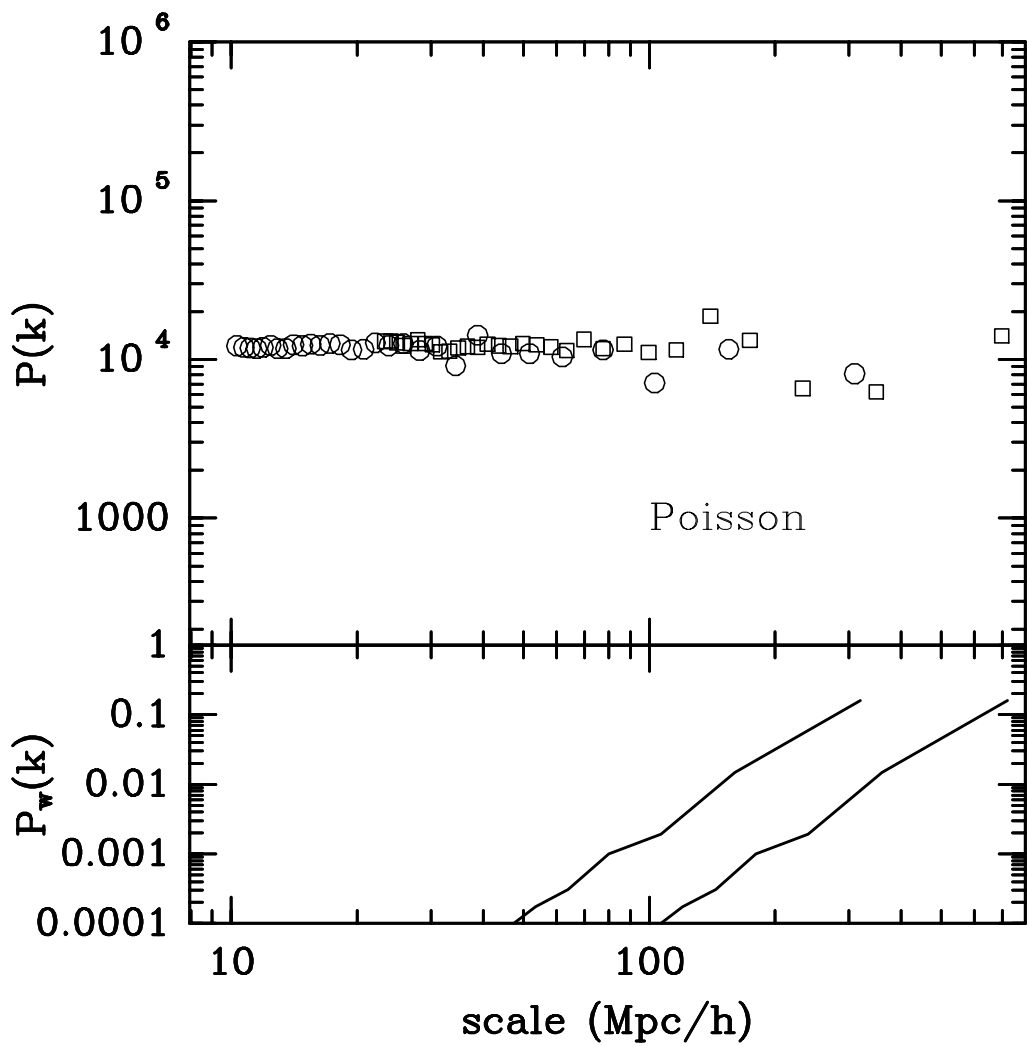
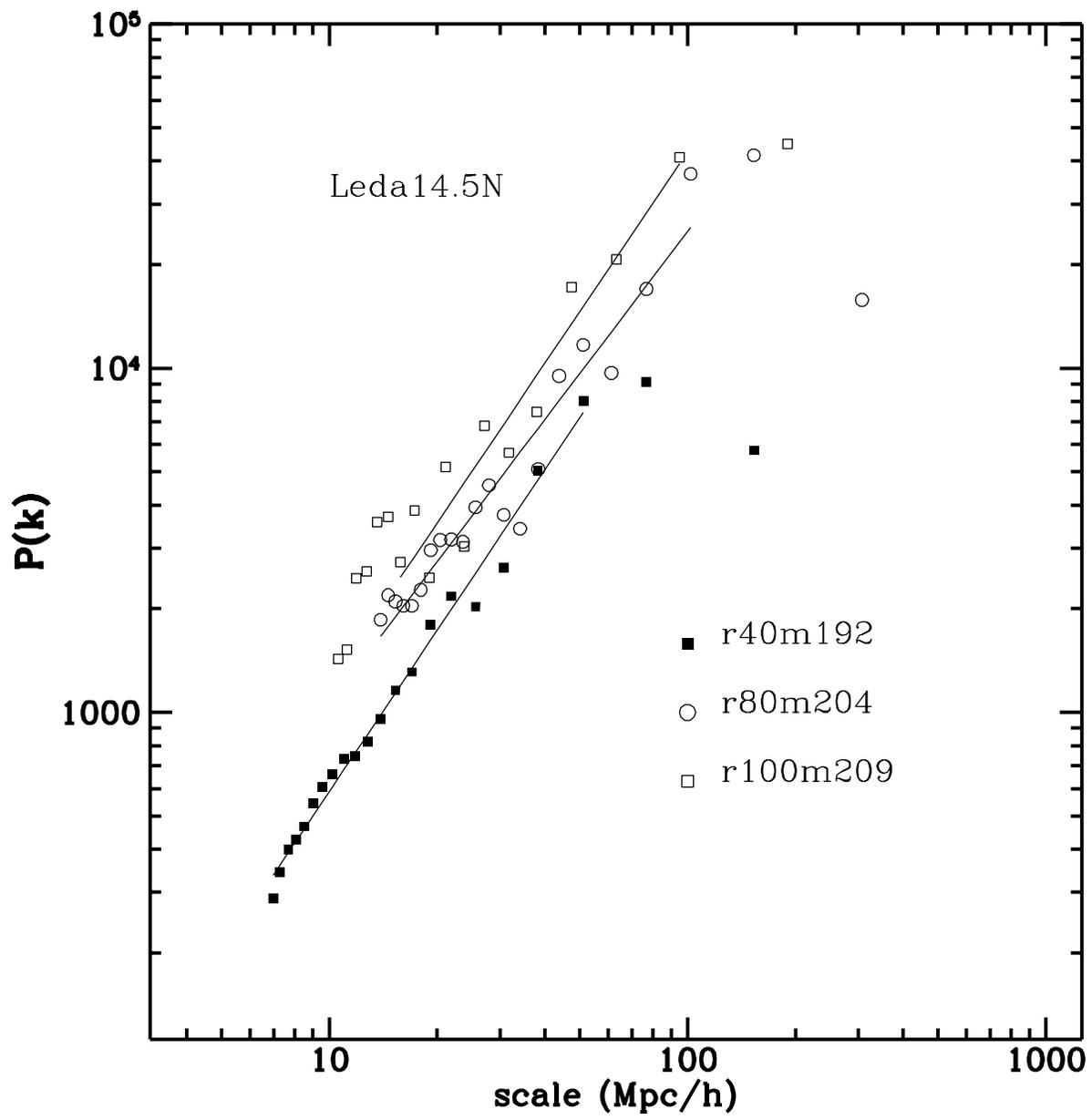
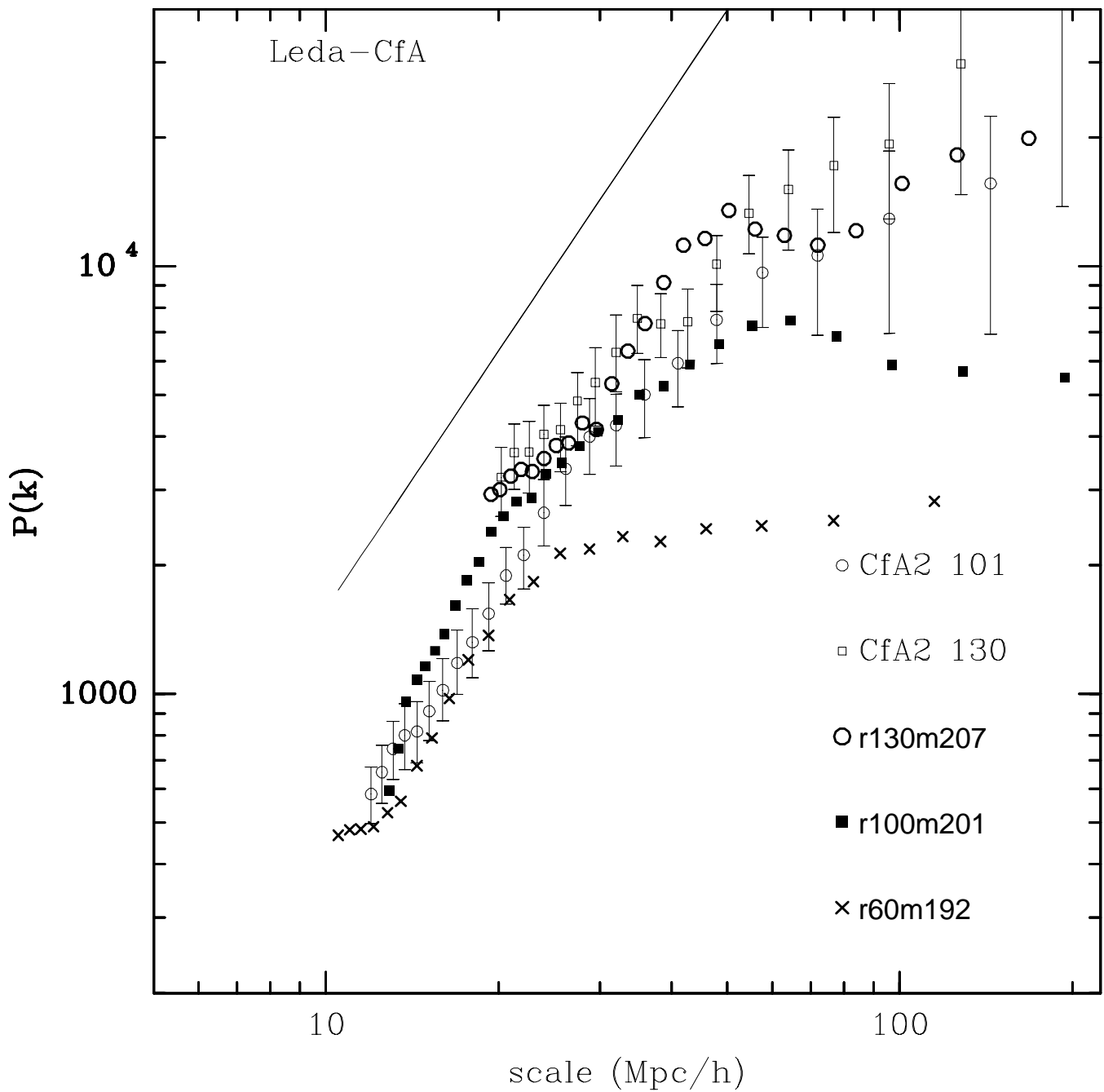
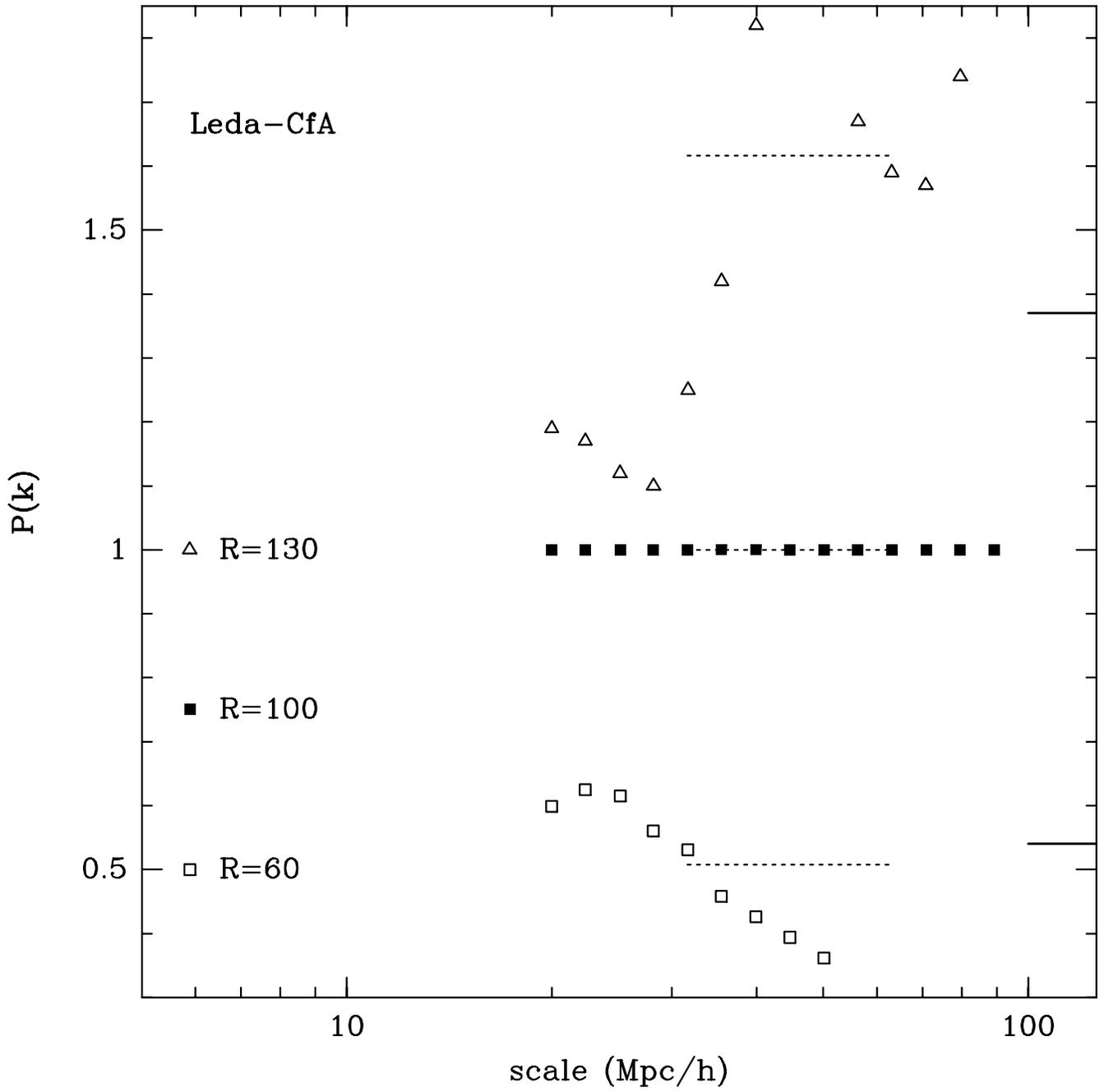
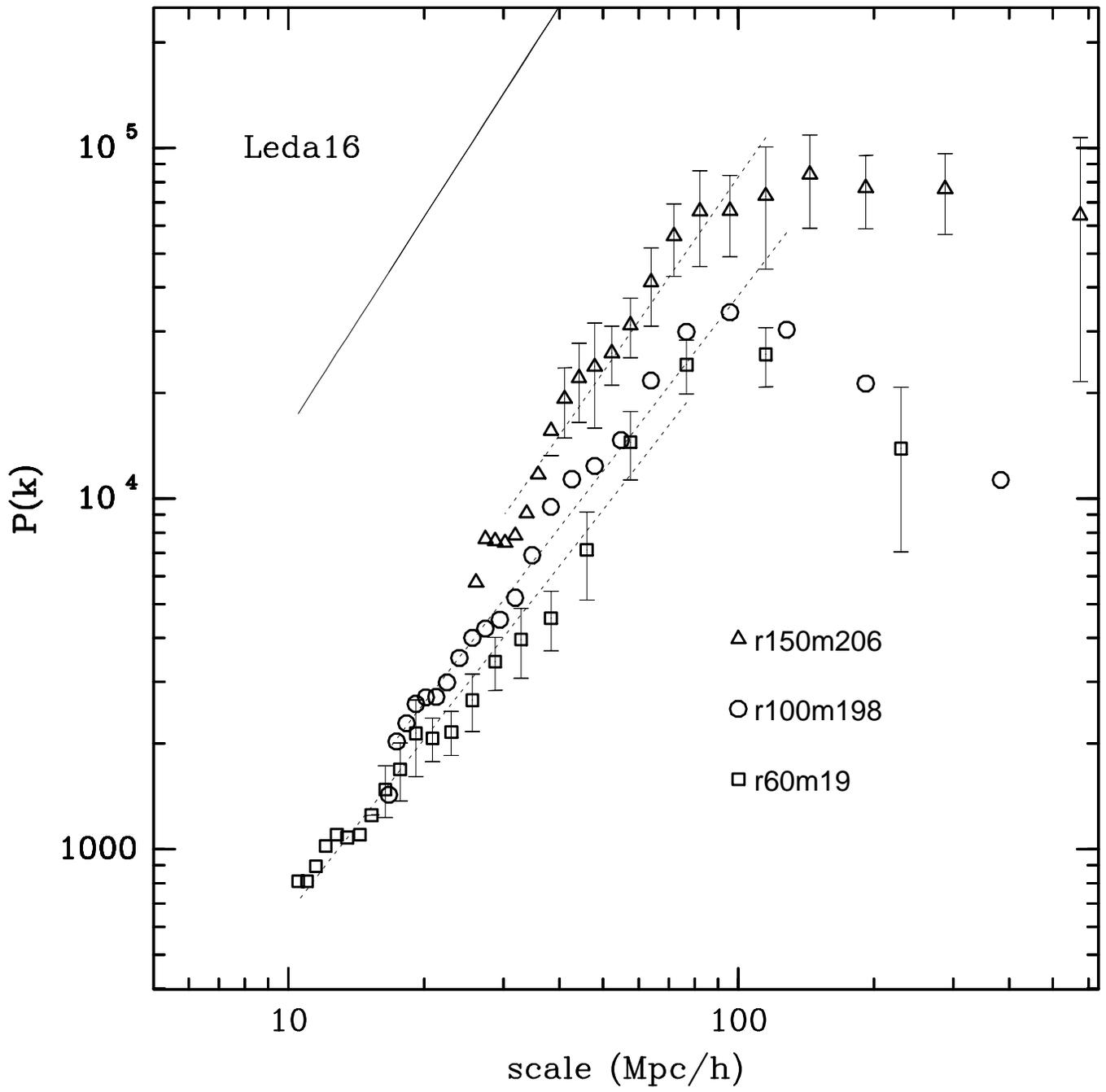
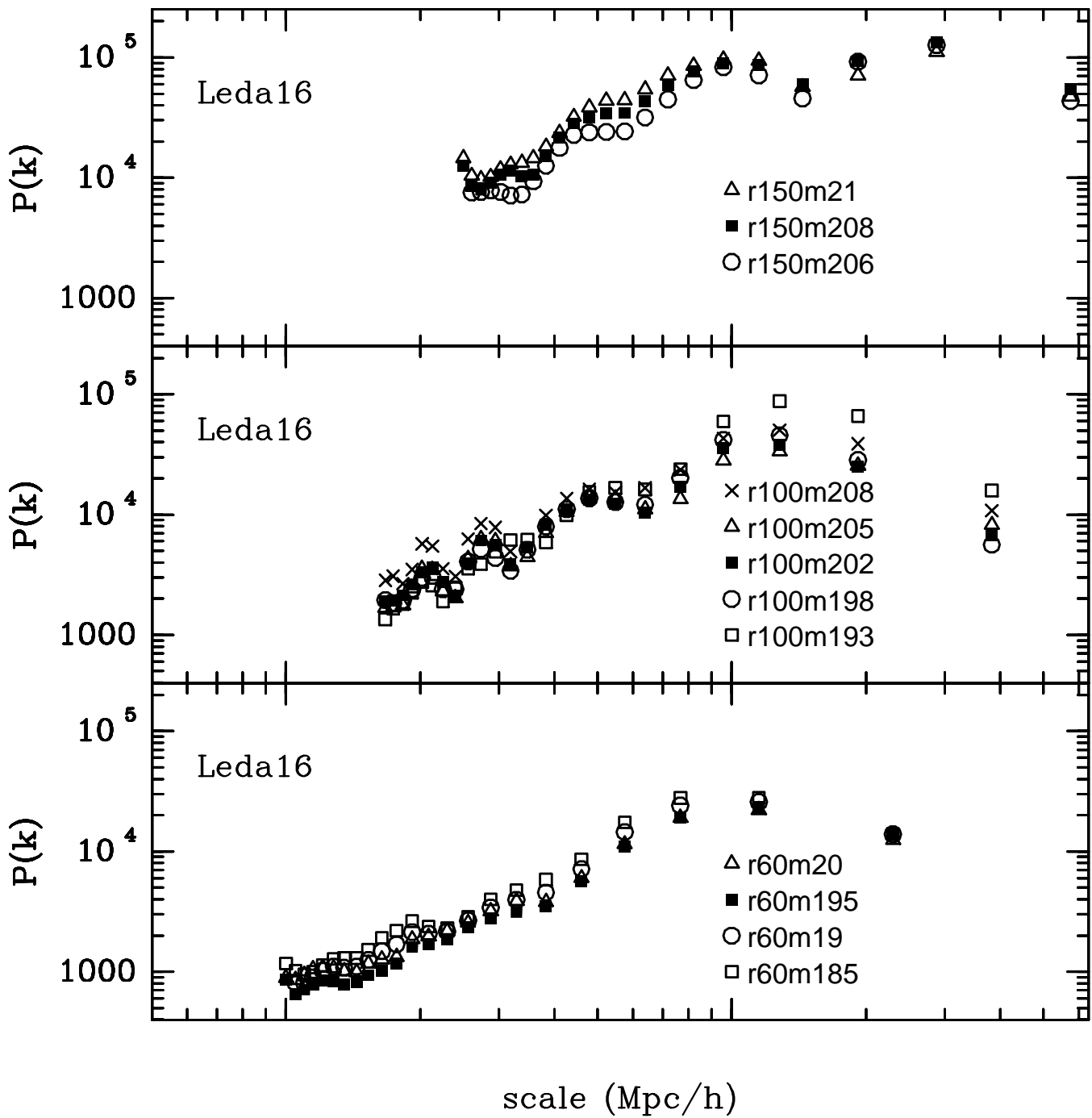
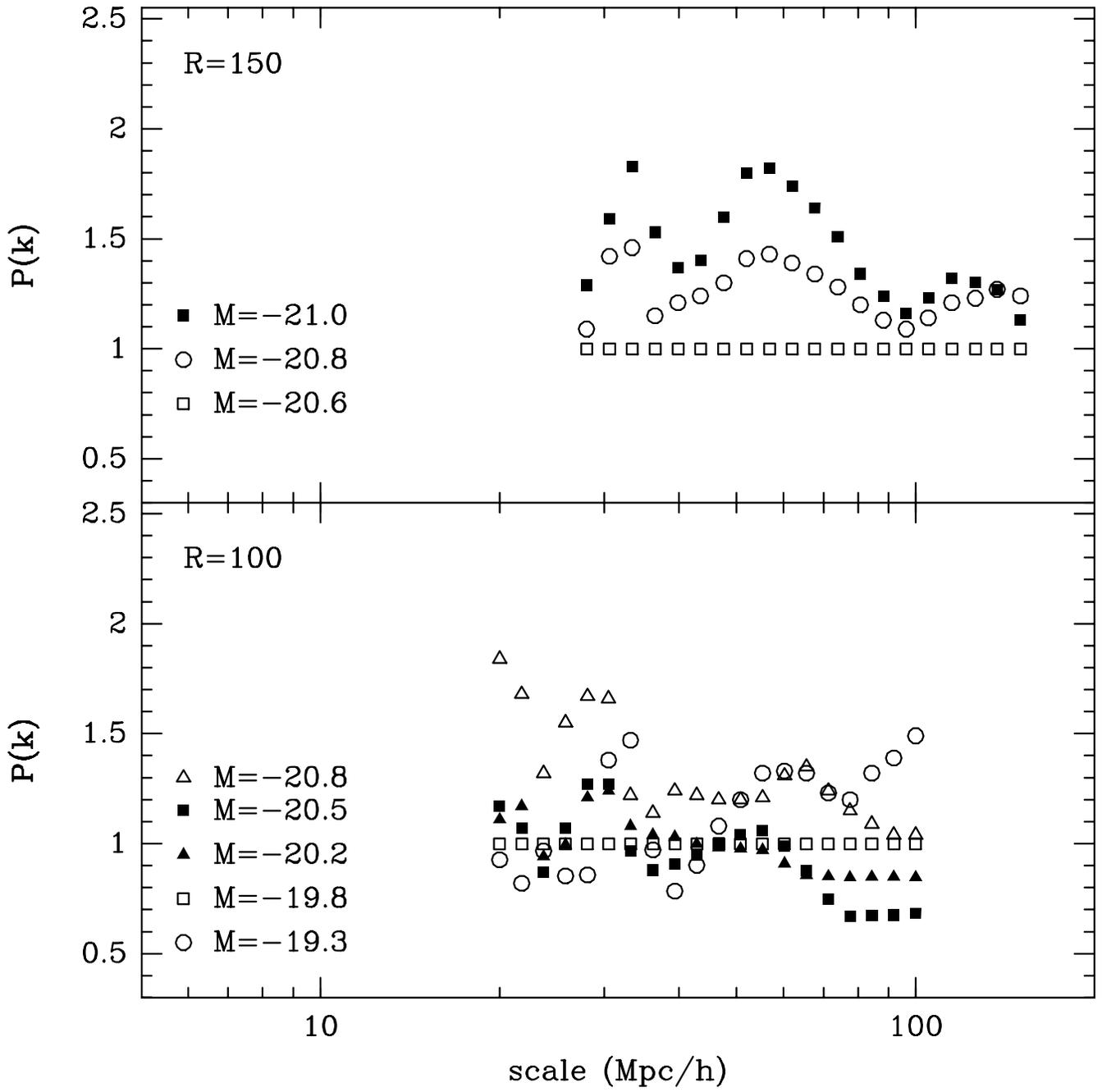
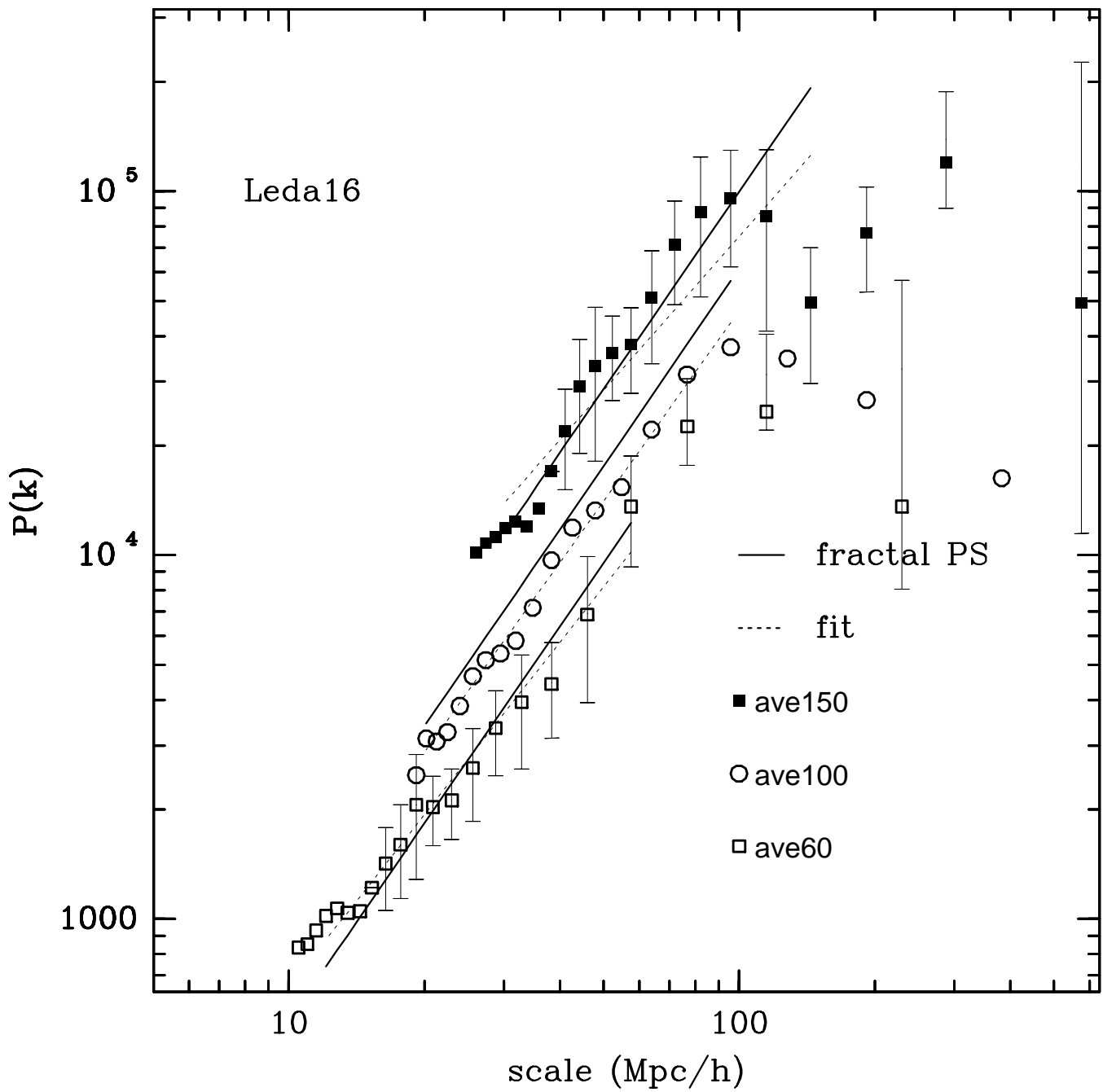
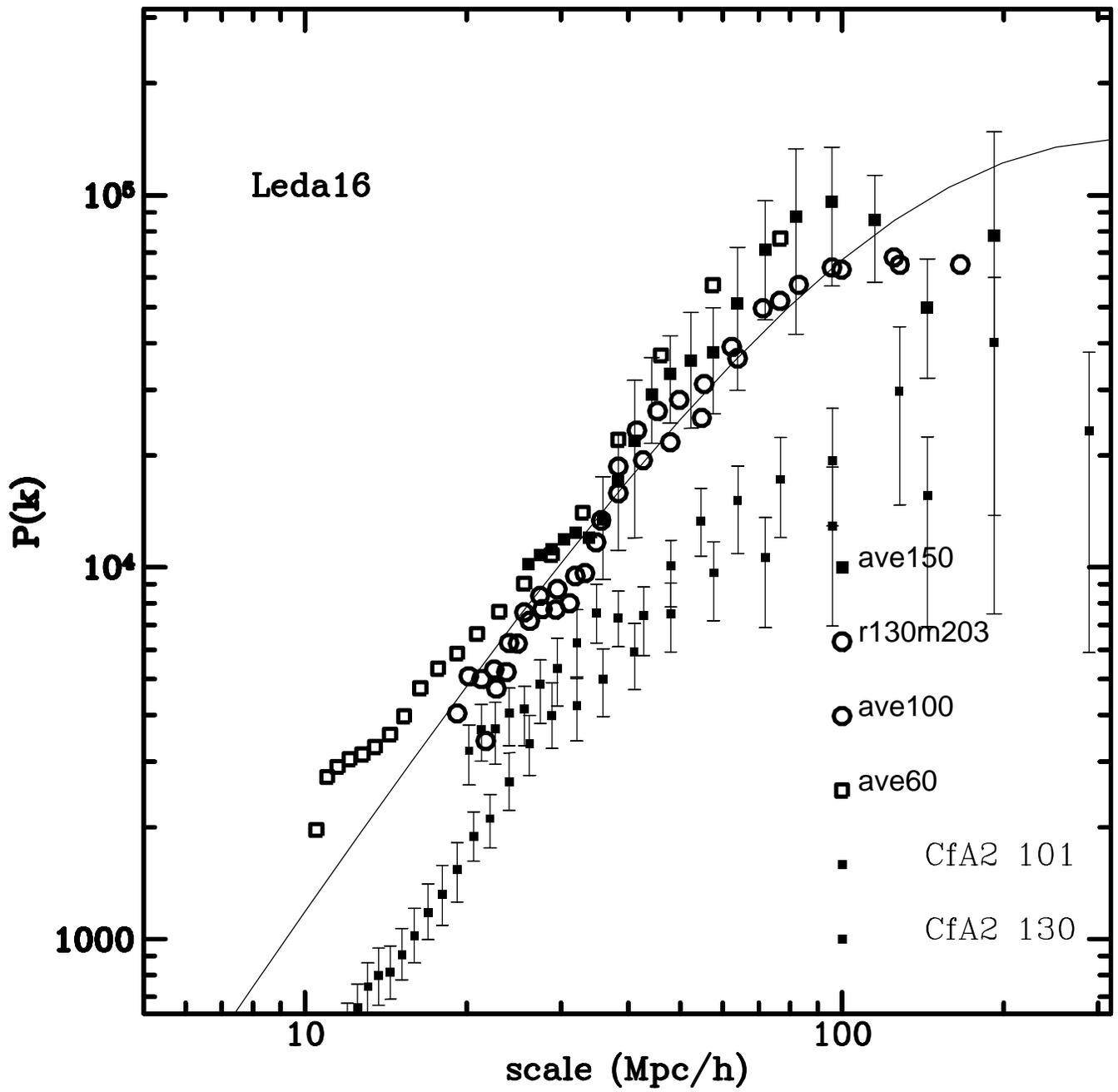
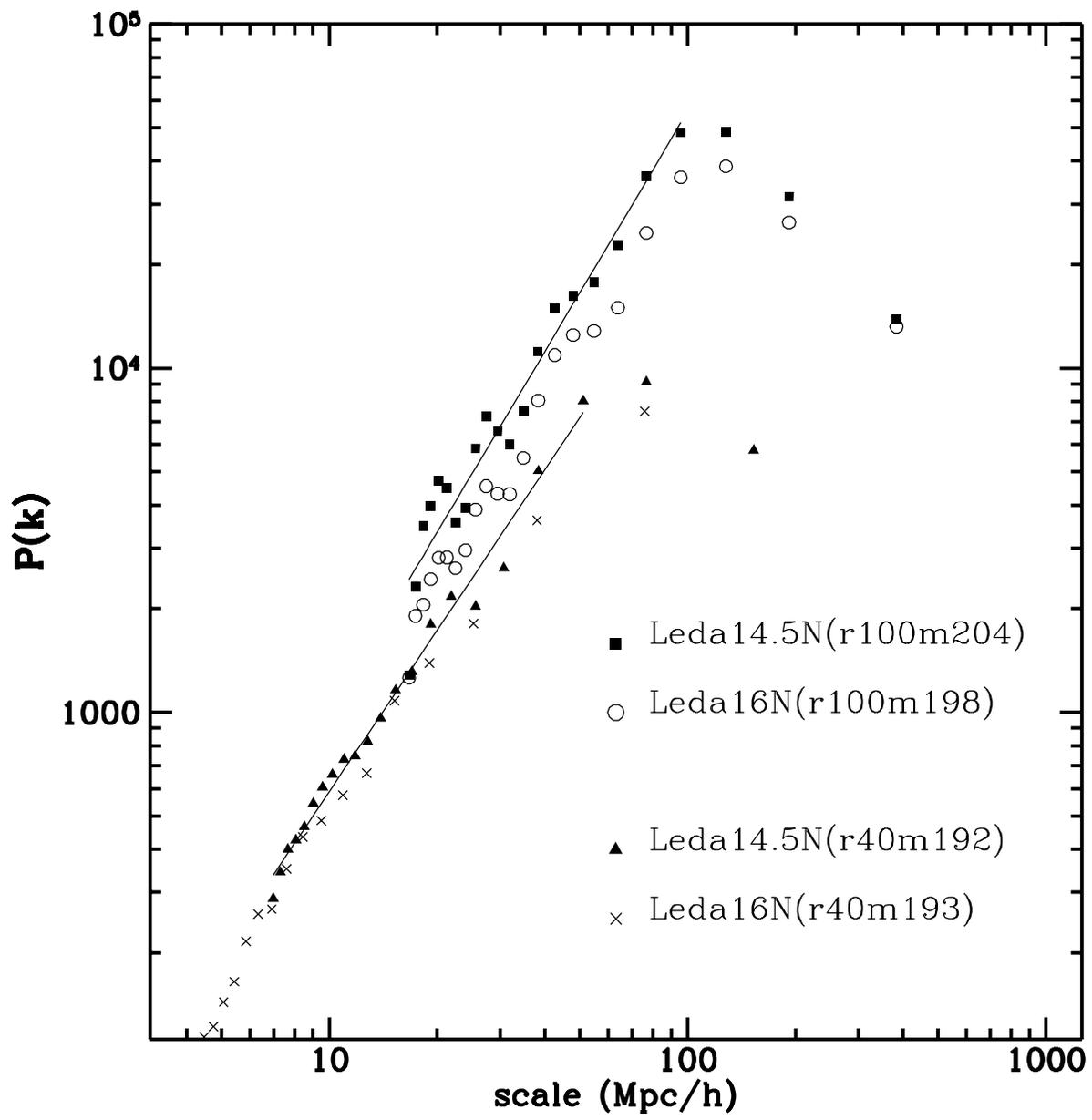
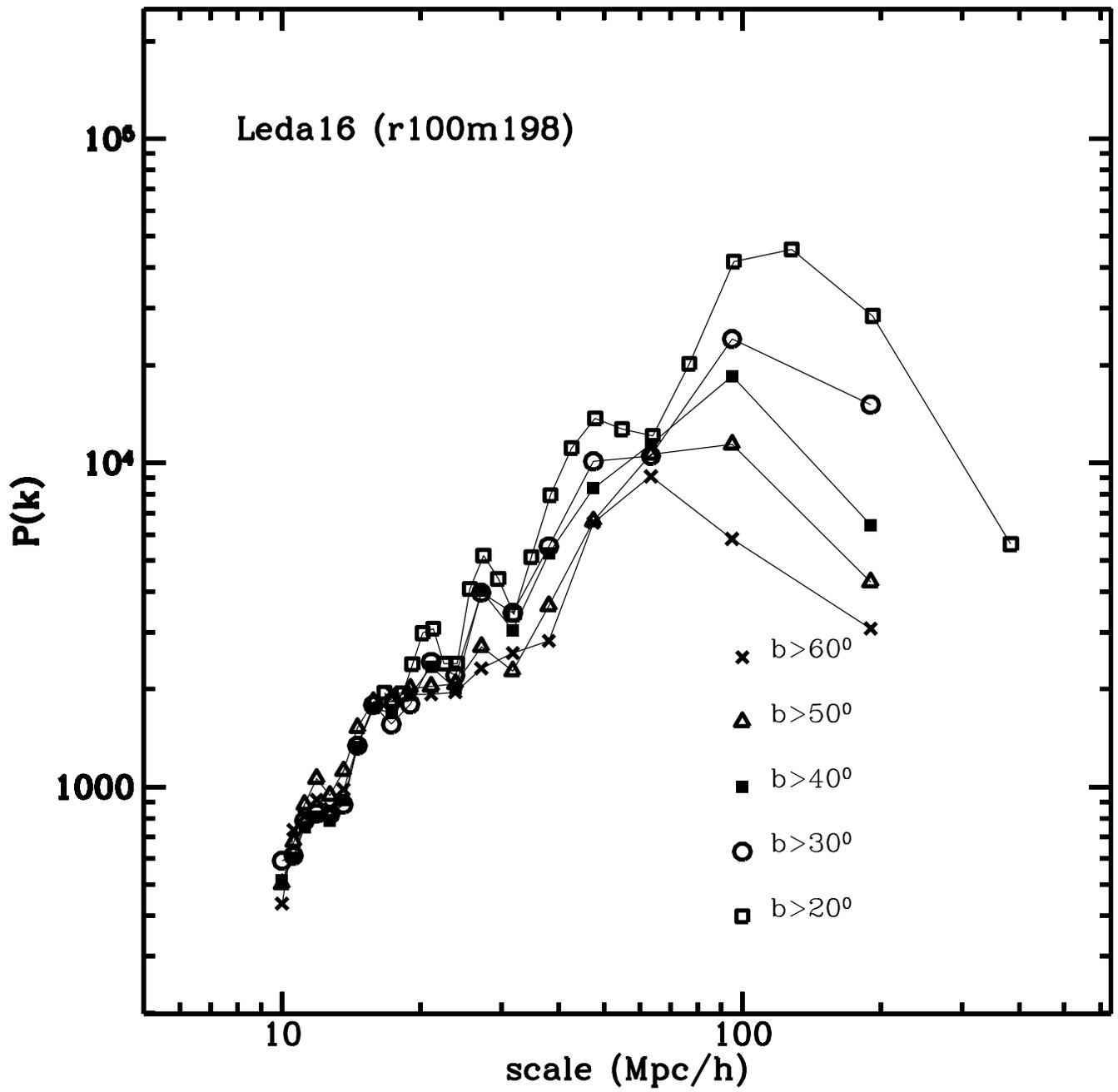
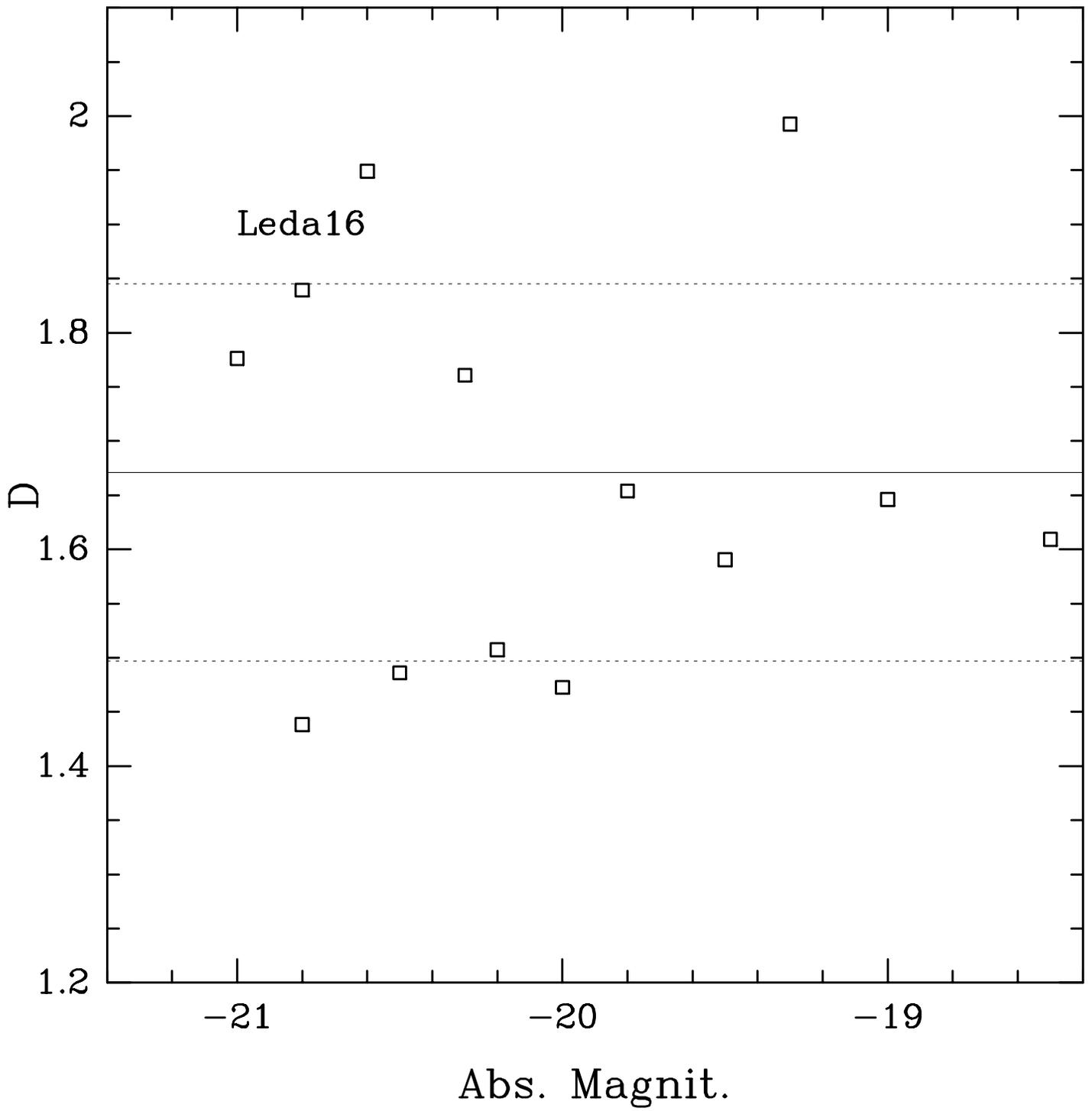
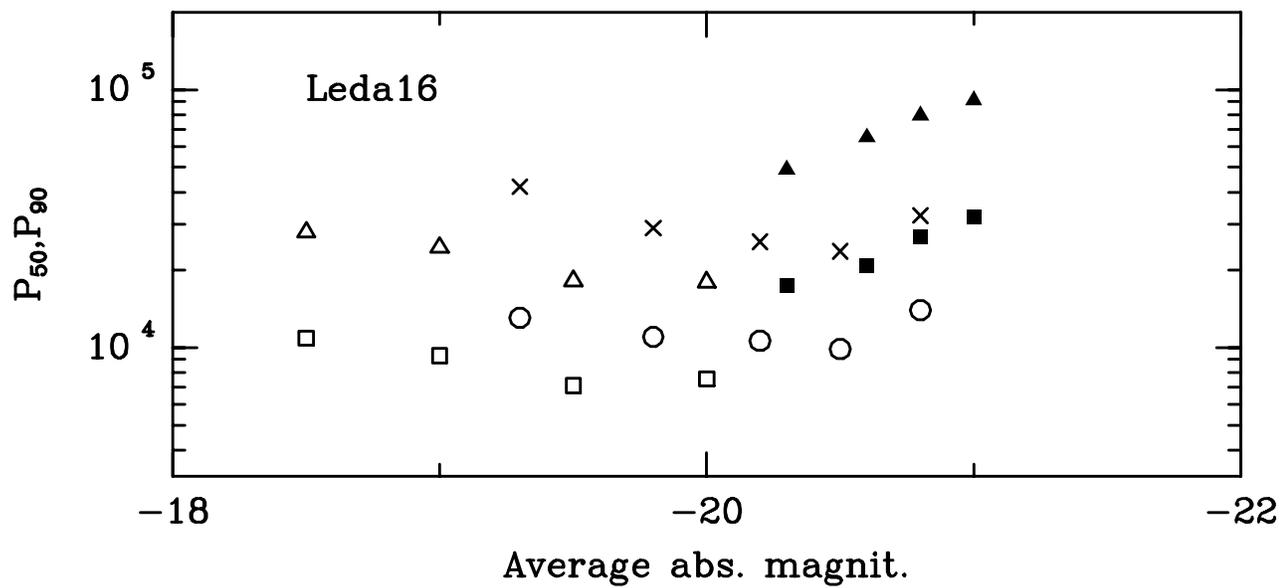
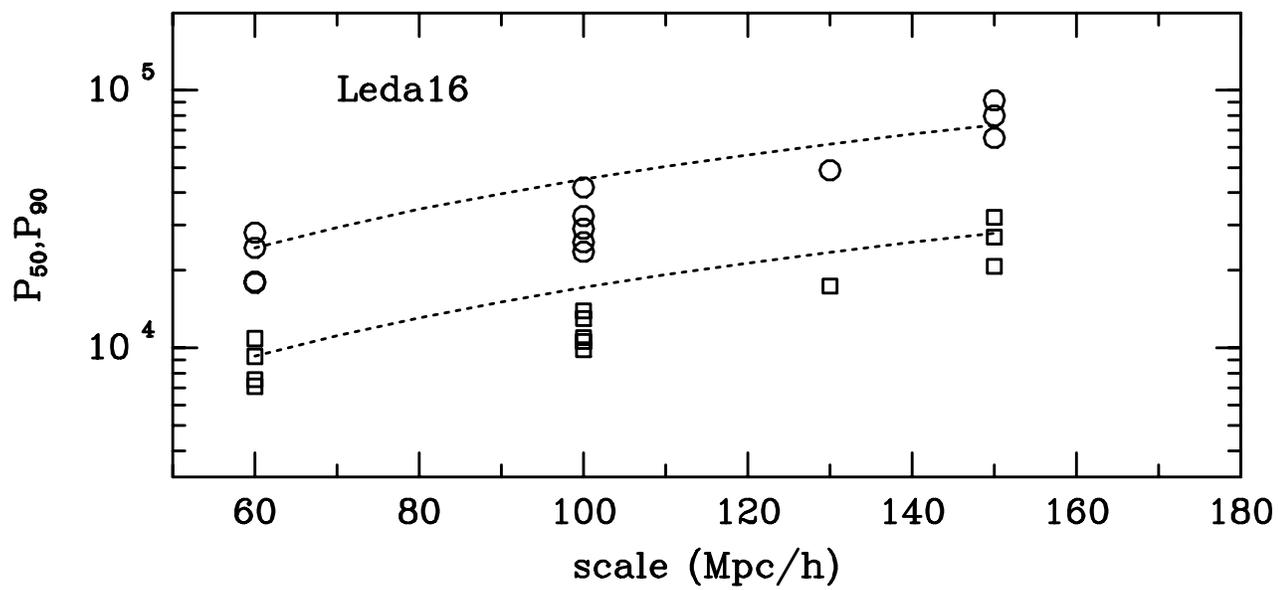
\end{document}